\documentclass[pdftex,twocolumn,epjc3]{svjour3}          

\RequirePackage[T1]{fontenc}

\smartqed  % flush right qed marks, e.g. at end of proof

\RequirePackage{graphicx}
\RequirePackage{mathptmx}      % use Times fonts if available on your TeX system
\RequirePackage[numbers,sort&compress]{natbib}
\RequirePackage[colorlinks,citecolor=blue,urlcolor=blue,linkcolor=blue]{hyperref}
\usepackage{amssymb}
\usepackage[usenames,dvipsnames]{xcolor}
\usepackage[normalem]{ulem}

\definecolor{forestgreen}{rgb}{0.13, 0.55, 0.13}
\definecolor{cssgreen}{rgb}{0.07, 0.53, 0.03}

\newcommand{\srzero}[0]{science run~0}
\newcommand{\monoenergetic}[0]{mono-energetic}

\journalname{Eur.~Phys.~J.~C}

\begin{document}

%%%%%%%%%%%%%%%%%%%%%%%%%%%%%%%%%%%%%%%%%%%%%%%%%%%%%%%%%%%%%%%%%%%%%%%%%%%%%%%%
%%%%%%%%%%%%%%%%%%%%%%%%%%%%%%%%%%%%%%%%%%%%%%%%%%%%%%%%%%%%%%%%%%%%%%%%%%%%%%%%
\title{The XENON1T Dark Matter Experiment}

%%%%%%%%%%%%%%%%%%%%%%%%%%%%%%%%%%%%%%%%%%%%%%%%%%%%%%%%%%%%%%%%%%%%%%%%%%%%%%%%
%%%%%%%%%%%%%%%%%%%%%%%%%%%%%%%%%%%%%%%%%%%%%%%%%%%%%%%%%%%%%%%%%%%%%%%%%%%%%%%%
\author{E.~Aprile\thanksref{add::columbia}
\and  J.~Aalbers\thanksref{add::nikhef}
\and  F.~Agostini\thanksref{add::lngs,add::bologna}
\and  M.~Alfonsi\thanksref{add::mainz}
\and  F.~D.~Amaro\thanksref{add::coimbra}
\and  M.~Anthony\thanksref{add::columbia}
\and  B.~Antunes\thanksref{add::coimbra}
\and  F.~Arneodo\thanksref{add::nyuad}
\and  M.~Balata\thanksref{add::lngs}
\and  P.~Barrow\thanksref{add::zurich}
% ...
 (XENON Collaboration\thanksref{e4}).
}

\thankstext{bern}{Also at Albert Einstein Center for Fundamental Physics, University of Bern, 3012 Bern, Switzerland}
\thankstext{wallenberg}{Wallenberg Academy Fellow}
\thankstext{coimbra}{Also at Coimbra Engineering Institute, Coimbra, Portugal}
\thankstext{e1}{\tt colijn@nikhef.nl}
\thankstext{e2}{\tt mmessina@astro.columbia.edu}
\thankstext{e3}{\tt marc.schumann@physik.uni-freiburg.de}
\thankstext{e4}{\tt xenon@lngs.infn.it}

\institute{Physics Department, Columbia University, New York, NY 10027, USA\label{add::columbia}
          \and
Nikhef and the University of Amsterdam, Science Park, 1098XG Amsterdam, Netherlands\label{add::nikhef}
          \and
INFN-Laboratori Nazionali del Gran Sasso and Gran Sasso Science Institute, 67100 L'Aquila, Italy\label{add::lngs}
          \and
Department of Physics and Astrophysics, University of Bologna and INFN-Bologna, 40126 Bologna, Italy\label{add::bologna}
          \and
Institut f\"{u}r Physik \& Exzellenzcluster PRISMA, Johannes Gutenberg-Universit\"{a}t Mainz, 55099 Mainz, Germany\label{add::mainz}
          \and
LIBPhys, Department of Physics, University of Coimbra, 3004-516 Coimbra, Portugal\label{add::coimbra}
          \and
New York University Abu Dhabi, Abu Dhabi, United Arab Emirates\label{add::nyuad}
          \and
Physik Institut, University of Zurich, 8057  Zurich, Switzerland\label{add::zurich}
          \and
Oskar Klein Centre, Department of Physics, Stockholm University, AlbaNova, Stockholm SE-10691, Sweden\label{add::stockholm}
          \and
Department of Physics, Applied Physics and Astronomy, Rensselaer Polytechnic Institute, Troy, NY 12180, USA\label{add::rpi}
          \and
Department of Particle Physics and Astrophysics, Weizmann Institute of Science, Rehovot 7610001, Israel\label{add::wis}
          \and
Max-Planck-Institut f\"{u}r Kernphysik, 69117 Heidelberg, Germany\label{add::heidelberg}
          \and
Physikalisches Institut, Universit\"{a}t Freiburg, 79104 Freiburg, Germany\label{add::freiburg}
          \and
Department of Physics and Astronomy, Purdue University, West Lafayette, IN 47907, USA\label{add::purdue}
          \and
SUBATECH, IMT Atlantique, CNRS/IN2P3, Universit\'{e} de Nantes, Nantes 44307, France\label{add::subatech}
          \and
Department of Physics, University of California, San Diego, CA 92093, USA\label{add::ucsd}
          \and
Institut f\"ur Kernphysik, Westf\"{a}lische Wilhelms-Universit\"{a}t M\"{u}nster, 48149 M\"{u}nster, Germany\label{add::munster}
          \and
INFN-Torino and Osservatorio Astrofisico di Torino, 10125 Torino, Italy\label{add::torino}
          \and
Department of Physics \& Kavli Institute of Cosmological Physics, University of Chicago, Chicago, IL 60637, USA\label{add::chicago}
          \and
Department of Physics and Astronomy, Rice University, Houston, TX 77005, USA\label{add::rice}
          \and
LPNHE, Universit\'e Pierre et Marie Curie, Universit\'{e} Paris Diderot, CNRS/IN2P3, Paris 75252, France\label{add::paris}
          \and
Physics \& Astronomy Department, University of California, Los Angeles, CA 90095, USA\label{add::ucla}
}

\date{}

\maketitle

\sloppy

%%%%%%%%%%%%%%%%%%%%%%%%%%%%%%%%%%%%%%%%%%%%%%%%%%%%%%%%%%%%%%%%%%%%%%
%% Abstract
%%%%%%%%%%%%%%%%%%%%%%%%%%%%%%%%%%%%%%%%%%%%%%%%%%%%%%%%%%%%%%%%%%%%%%

\begin{abstract}
The XENON1T experiment at the Laboratori Na\-zionali del Gran Sasso (LNGS) is the first WIMP dark matter detector operating with a liquid xenon target mass above the ton-scale. Out of its 3.2\,t liquid xenon inventory, 2.0\,t constitute the active target of the dual-phase time projection chamber. The scintillation and ionization signals from particle interactions are detected with low-background photomultipliers. This article describes the XENON1T instrument and its subsystems as well as strategies to achieve an unprecedented low background level. First results on the detector response and the performance of the subsystems are also presented.\end{abstract}

%%%%%%%%%%%%%%%%%%%%%%%%%%%%%%%%%%%%%%%%%%%%%%%%%%%%%%%%%%%%%%%%%%%%%%
%%%%%%%%%%%%%%%%%%%%%%%%%%%%%%%%%%%%%%%%%%%%%%%%%%%%%%%%%%%%%%%%%%%%%%
%%%%%%%%%%%%%%%%%%%%%%%%%%%%%%%%%%%%%%%%%%%%%%%%%%%%%%%%%%%%%%%%%%%%%%
\section{Introduction}
\label{sec::introduction}

The fact that dark matter exists, as evidenced by a large number of indirect observations in astronomy and cosmology~\cite{ref::dm_evidence}, is seen as a striking indication that there must be new physics beyond the Standard Model (BSM) of particle physics. The postulated dark matter particle has not been directly observed yet, and theoretical predictions about its mass, couplings and production mechanisms span a large parameter space~\cite{ref::dm_predictions}. A well-motivated candidate, which arises naturally in several BSM models, is the weakly interacting massive particle (WIMP)~\cite{ref::wimp}. It might be directly detectable in sensitive Earth-based detectors, as it is expected to scatter off the detector's target nuclei. Most models predict an exponentially falling nuclear recoil spectrum, with mean energies of a few~keV~\cite{ref::directdetection}. 

The XENON dark matter project aims at the detection of WIMP dark matter with dual-phase time projection chambers filled with a liquid xenon (LXe) target. The first WIMP search conducted with XENON10~\cite{ref::xe10_si,ref::xe10_instr} featured a target mass of 14\,kg (25\,kg total). It was followed by XENON100 (62\,kg target, 161\,kg total mass)~\cite{ref::xe100_instr}, which published competitive results on spin-independent~\cite{ref::xe100_si,ref::xe100_sicomb}, spin-dependent~\cite{ref::xe100_sd} and other WIMP-nucleon interactions~\cite{ref::xe100_inelastic,Aprile:2017kek,Aprile:2017aas}, axions and axion-like particles~\cite{ref::xe100_axions}, and challenged the interpretation of the DAMA/LIBRA signal as being due to leptophilic dark matter interacting with atomic electrons~\cite{ref::xe100_ac,ref::xe100_accomb,ref::xenon100_dc}. 

The XENON1T experiment described in this article is located underground in Hall~B of the Laboratori Nazionali del Gran Sasso (LNGS), Italy, at a depth of 3600\,meter water equivalent. With its large target mass of 2.0\,t (2000\,kg) it aims at probing spin-independent WIMP-nucleon scattering cross sections of $1.6 \times 10^{-47}$\,cm$^2$ at a WIMP mass of $m_\chi = 50$\,GeV/$c^2$, with an exposure of 2.0\,t\,$\times$\,y~\cite{ref::xe1t_reach}. At low WIMP masses, the sensitivity approaches the predicted ``neutrino floor''~\cite{ref::billard} caused by background events from the coherent scattering of solar $^8$B neutrinos off the xenon nuclei. The first dark matter search results from XENON1T (from ``\srzero'') probe spin-inde\-pen\-dent WIMP-nucleon cross sections below~$1\times 10^{-46}$\,cm$^2$ for the first time~\cite{ref::xe1t_sr0}.

Most XENON1T subsystems were designed such that they can also support a significantly larger dark matter detector, with a target of $\sim$6\,t. This phase of the project, XENONnT, is being prepared during XENON1T data taking to allow for a rapid exchange of the larger instrument after the science goals of XENON1T will have been reached. XENONnT aims at improving the spin-independent WIMP sensitivity by another order of magnitude compared to XENON1T~\cite{ref::xe1t_reach}.

The article is structured as follows: in section~\ref{sec::xe1t}, the XENON1T experiment with all its subsystems is introduced in detail. Section~\ref{sec::commissioning} presents selected results from detector commissioning and from \srzero, and section~\ref{sec::outlook} provides an outlook.

%%%%%%%%%%%%%%%%%%%%%%%%%%%%%%%%%%%%%%%%%%%%%%%%%%%%%%%%%%%%%%%%%%%%%%
%%%%%%%%%%%%%%%%%%%%%%%%%%%%%%%%%%%%%%%%%%%%%%%%%%%%%%%%%%%%%%%%%%%%%%
%%%%%%%%%%%%%%%%%%%%%%%%%%%%%%%%%%%%%%%%%%%%%%%%%%%%%%%%%%%%%%%%%%%%%%
\section{The XENON1T Experiment}
\label{sec::xe1t}

This section describes the XENON1T detector. The dual-phase TPC (section~\ref{sec::tpc}) is installed inside a double-walled vacuum cryostat (section~\ref{sec::cryostat}) in the center of a large water tank. The tank serves as passive shield as well as a Cherenkov muon veto (section~\ref{sec::muonveto}). A three-floor building accommodates all auxiliary systems. These include the systems to cool (section~\ref{sec::cryostat}), store, and purify the xenon gas (section~\ref{sec::xenon}), the cryogenic distillation column for krypton removal (section~\ref{sec::kr}), the data acquisition system (section~\ref{sec::daq}) as well as the control and monitoring systems for the entire experiment (section~\ref{sec::slowcontrol}). The TPC calibration systems are installed on the purification system as well as on the top of the water shield (section~\ref{sec::calibration}). 

%%%%%%%%%%%%%%%%%%%%%%%%%%%%%%%%%%%%%%%%%%%%%%%%%%%%%%%%%%%%%%%%%%%%%%
%%%%%%%%%%%%%%%%%%%%%%%%%%%%%%%%%%%%%%%%%%%%%%%%%%%%%%%%%%%%%%%%%%%%%%
%%%%%%%%%%%%%%%%%%%%%%%%%%%%%%%%%%%%%%%%%%%%%%%%%%%%%%%%%%%%%%%%%%%%%%
\subsection{Time Projection Chamber}
\label{sec::tpc}

This section describes the working principle and design of the XENON1T TPC, the photosensors (photomultipliers, PMTs) to register particle interactions, and the systems that handle the $\sim$3.2\,t of xenon in liquid and gaseous form. All materials and components constituting the TPC were selected for a low intrinsic radioactivity (see section~\ref{sec::materials}). 

%%%%%%%%%%%%%%%%%%%%%%%%%%%%%%%%%%%%%%%%%%%%%%%%%%%%%%%%%%%%%%%%%%%%%%
\subsubsection{Working Principle}
\label{sec::tpcdesign}

\begin{figure}[t!]
\includegraphics*[width=0.48\textwidth]{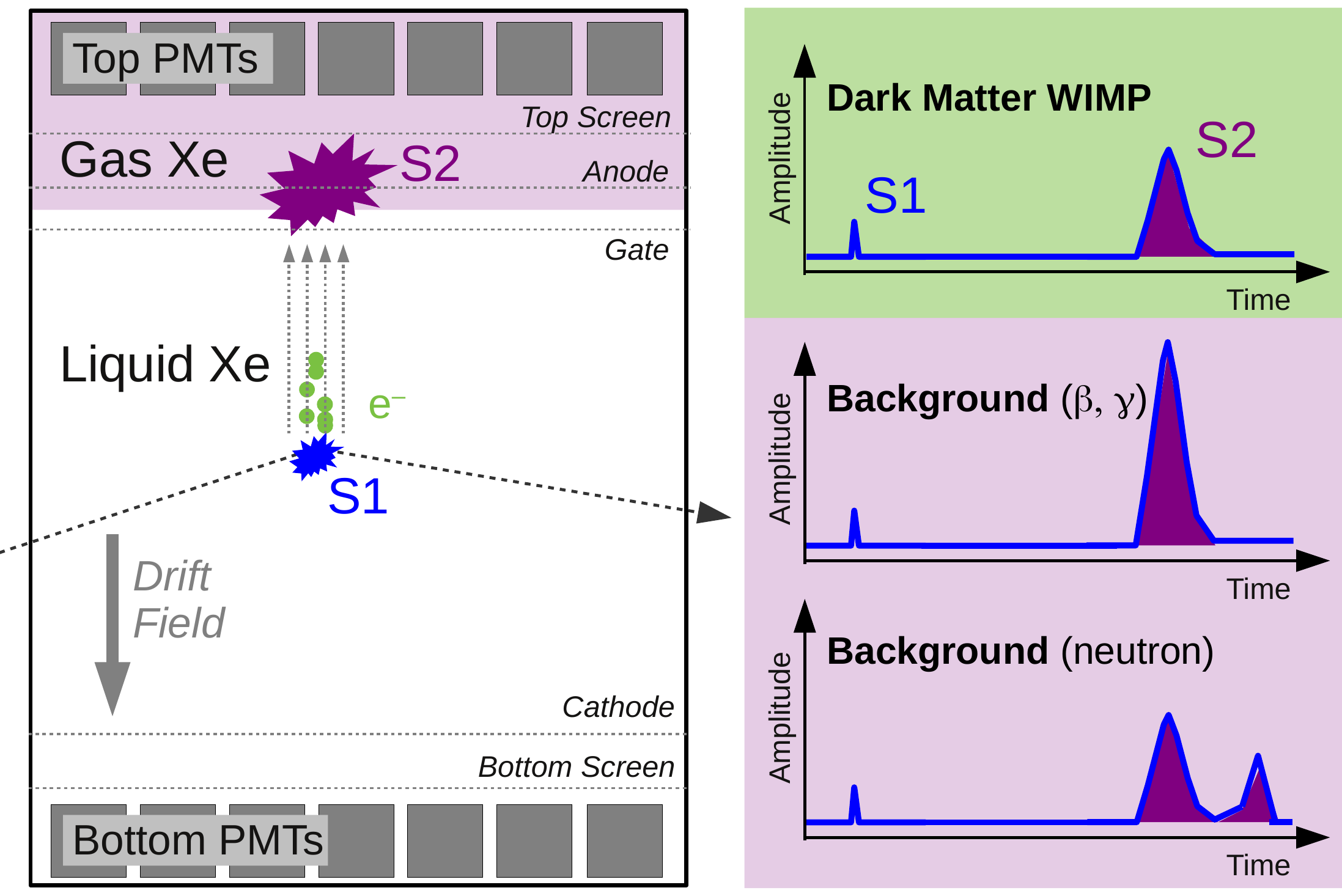}
\caption{Working principle of a dual-phase LXe TPC: The prompt scintillation light (S1) generated in the LXe is recorded by PMTs installed above and below the target. The same PMTs also measure the delayed secondary-light signal S2, which is created by proportional scintillation induced by ionization electrons (e$^-$) in the gas phase. A set of TPC electrodes is used to establish the required electric fields. The interaction position can be reconstructed in 3-dimensions by means of the S2-signal pattern observed by the top PMTs (lateral) and the time difference between S1 and S2 (depth). Background events are rejected by the charge to light (S2/S1) ratio and the scatter multiplicity (number of S2~signals), as indicated on the panels on the right.\label{fig::dualphase}}
\end{figure}

Dual-phase TPCs~\cite{ref::doublephase} were first used for WIMP dark matter searches by the ZEPLIN-II~\cite{ref::zeplin} and XENON10~\cite{ref::xe10_si} collaborations and are now successfully employed by number of experiments~\cite{ref::2phase_chepel,ref::2phase_schumann}. The working principle is illustrated in figure~\ref{fig::dualphase}: particles entering a cylindrical LXe target can scatter off xenon nuclei (in case of WIMPs or neutrons) or can interact with atomic electrons ($\gamma$ rays and $\beta$ electrons), generating nuclear recoils or electronic recoils, respectively. The recoils excite and ionize the LXe; some energy is lost to heat. The partition into the different energy-loss channels depends on the recoil type and energy and can therefore be used to distinguish a WIMP signal from electronic recoil backgrounds, provided that the resulting excitation and ionization signals can be measured independently~\cite{ref::discrimination}. The Xe$^*_2$ excimers, that are eventually formed, de-excite under the emission of 178\,nm scintillation light. In dual-phase TPCs, this light signal (S1) is observed by photosensors installed above and below the target. An electric field across the target, established between the negatively biased cathode at the bottom of the TPC and the gate electrode at ground potential at the top, is used to move the ionization electrons away from the interaction site, drifting them to the liquid-gas interface. A second field, generated between the gate and the positively-biased anode, extracts them into the gas phase and provides the electrons sufficient energy to excite and ionize the gas atoms. This generates a secondary scintillation signal (S2) which is proportional to the number of extracted electrons~\cite{ref::secondscint}. The position of the initial interaction, as well as the scatter multiplicity, can be reconstructed in 3-dimensions from the position and number of S2~signals observed by the top photosensors and the S1-S2~time difference. The ratio S2/S1 can be employed for electronic recoil background rejection, with typically $>$99.5\% discrimination at 50\% signal acceptance.

%%%%%%%%%%%%%%%%%%%%%%%%%%%%%%%%%%%%%%%%%%%%%%%%%%%%%
\subsubsection{XENON1T TPC}
\label{sec::xe1ttpc}

\begin{figure}[t!]
\includegraphics*[width=0.48\textwidth]{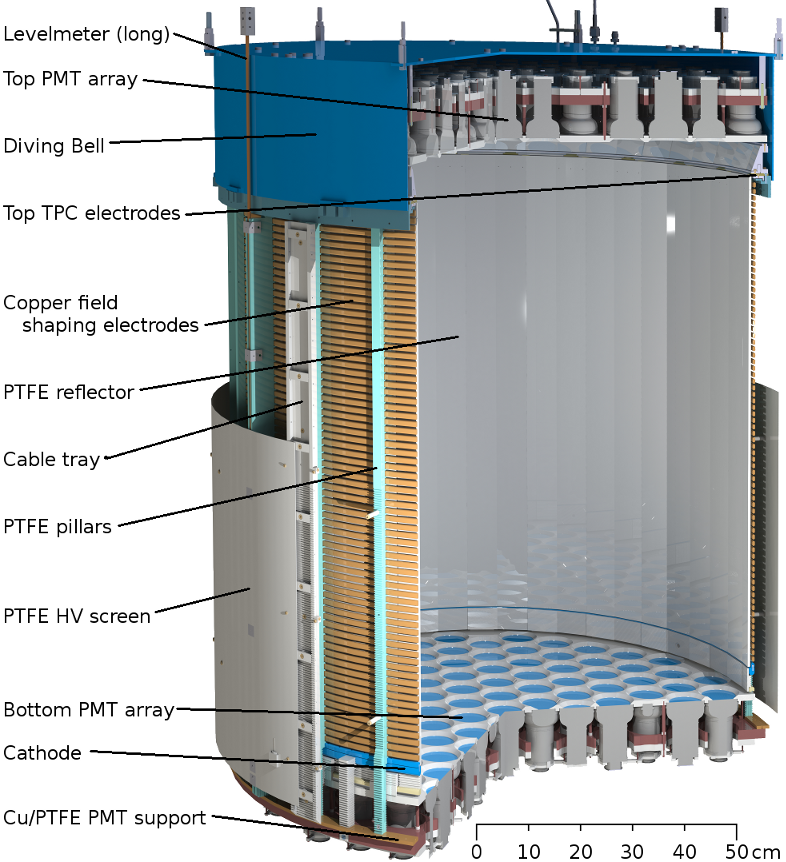}
\caption{Illustration of the XENON1T TPC. It is built from materials selected for their low radioactivity, e.g., OFHC copper, stainless steel and PTFE. The top and bottom PMT arrays are instrumented with 127~and 121~Hamamatsu R11410-21 PMTs, respectively.\label{fig::tpc_cad}}
\end{figure}

The cylindrical TPC of 97\,cm length and 96\,cm diameter contains an active LXe target of 2.0\,t, in which the light and the charge signals from each interaction can be detected, see figure~\ref{fig::tpc_cad}. It is enclosed by 24~interlocking and light-tight PTFE (polytetrafluoroethylene, Teflon) panels, whose surfaces were treated with diamond tools in order to optimize the reflectivity for vacuum ultraviolet (VUV) light~\cite{ref::ptfe_reflect}. Due to the rather large linear thermal expansion coefficient of PTFE, its length is reduced by about 1.5\% at the operation temperature of $-96^\circ$C. An interlocking design allows the radial dimension to remain constant while the vertical length is reduced. 

\begin{table*}[]
\caption{Specifications of the five TPC electrodes. The bottom screening electrode
features a single wire installed perpendicularly mid-way to all others to minimize elastic deformation of the frame. The last column indicates the vertical position of the electrodes inside the TPC. The distance between the top (bottom) screen to the top (bottom) PMTs is 11\,mm (12\,mm).}
\label{tab::meshes}
\centering
\small
\begin{tabular*}{\textwidth}{@{\extracolsep{\fill}}llllllr@{}}
\hline 
TPC Electrode & Type & Material & Wire Diameter & Pitch/Cell opening & Transparency & Position~~~~~~~\\ \hline
Top screen & hex etched & stainless steel & 178\,$\mu$m & 10.2\,mm & 96.5\% & 63\,mm\\
Anode & hex etched & stainless steel& 178\,$\mu$m & ~~3.5\,mm & 89.8\% & 5\,mm\\
Gate & hex etched & stainess steel& 127\,$\mu$m & ~~3.5\,mm &  92.7\% & 0\,mm\\
Cathode & parallel wires & Au plated stainless steel& 216\,$\mu$m & ~~7.75\,mm & 97.2\% & $-$969\,mm\\
Bottom screen & parallel wires & Au plated stainless steel& 216\,$\mu$m & ~~7.75\,mm & 97.2\% & $-$1017\,mm\\ \hline 
\end{tabular*}
\end{table*}

\begin{figure}[b!]
\includegraphics*[width=0.48\textwidth]{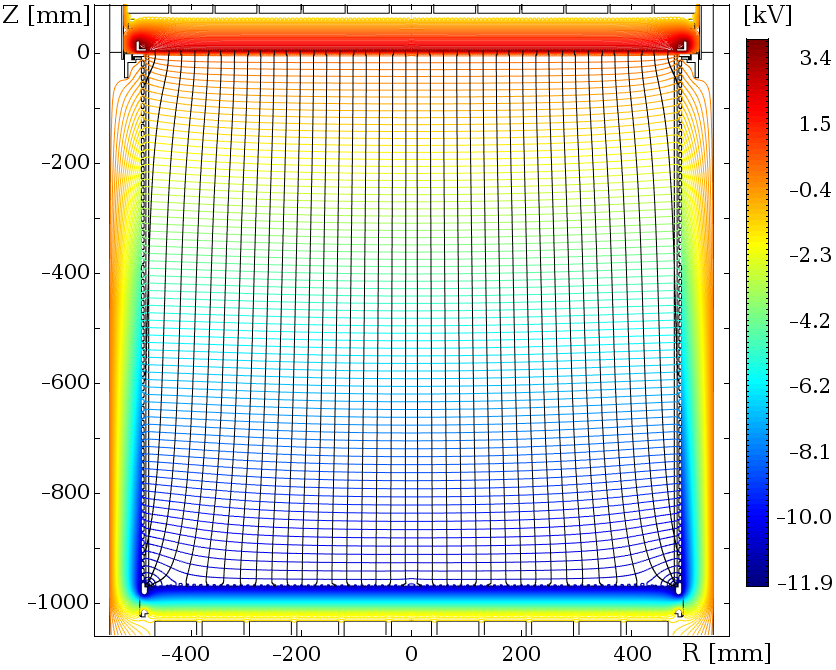}
\caption{Finite element simulation of the electric field configuration inside and outside of the TPC, separated by the gate and cathode as well as the field-shaping electrodes. The figure shows the field lines as well as the equipotential lines for cathode, gate and anode biased with $-$12\,kV, 0\,kV and $+$4\,kV, respectively, as realized during \srzero. \label{fig::tpc_efield}}
\end{figure}

To ensure drift field homogeneity, the TPC is surrounded by 74~field shaping electrodes with a cross section of $\sim$\,10\,$\times$\,5\,mm$^2$; they are made from low-radioac\-tivity oxygen-free high thermal conductivity (OFHC) copper. The electrodes are connected by two redundant chains of 5\,G$\Omega$ resistors; a 25\,G$\Omega$ resistor connects each chain to the cathode. The resistor settings, as well as the electrical transparency of the TPC electrodes (gate, anode and screening electrode on top, and cathode and screening electrode on bottom), were optimized by means of electrostatic simulations, using finite element (COMSOL Multiphysics~\cite{ref::comsol}) and boundary element methods (KEMField \cite{ref::kemfield}). The high-voltage configuration realized during \srzero~is shown in figure~\ref{fig::tpc_efield}. Most S1~light is detected by the photosensors below the target. The electrodes were thus designed for S1~light collection by optimizing the optical transparency of the gate, the cathode and the bottom screening electrodes. The details are summarized in table~\ref{tab::meshes}. The circular stainless-steel frames supporting the electrodes are electropolished and optimized for high-voltage operation. The etched meshes were spot-welded to the frames, while the single wires were pre-stretched on an external structure and fixed by wedging them between the upper and lower parts of the frames. Gold-plated wires were used to increase the work function of the metal.

The cathode is negatively biased using a Heinzinger~PNC\,150000-1\,NEG high-voltage supply via a custom-made high-voltage feedthrough. The latter consists of a conducting stainless-steel rod inside an ultra-high molecular weight (UHMW) polyethylene insulator, cryofitted into a 25.4\,mm diameter stainless-steel tube to make it vacuum tight. Before installation, the feedthrough was successfully tested to voltages exceeding $-$100\,kV. The total length of the feedthrough is about~1.5\,m, out of which 1.2\,m are located inside the cryostat. This ensures that the connection point to the PTFE insulated metal rod, which supplies the voltage to the cathode frame, is covered by LXe. The anode is positively biased by a CAEN~A1526P unit via a commercial Kapton-insu\-lated cable (Accuglass 26\,AWG, TYP22-15). The gate electrode is kept at ground potential and the screening electrodes can be biased to minimize the field in front of the PMT photocathodes.

\begin{figure}[b!]
\includegraphics*[width=0.48\textwidth]{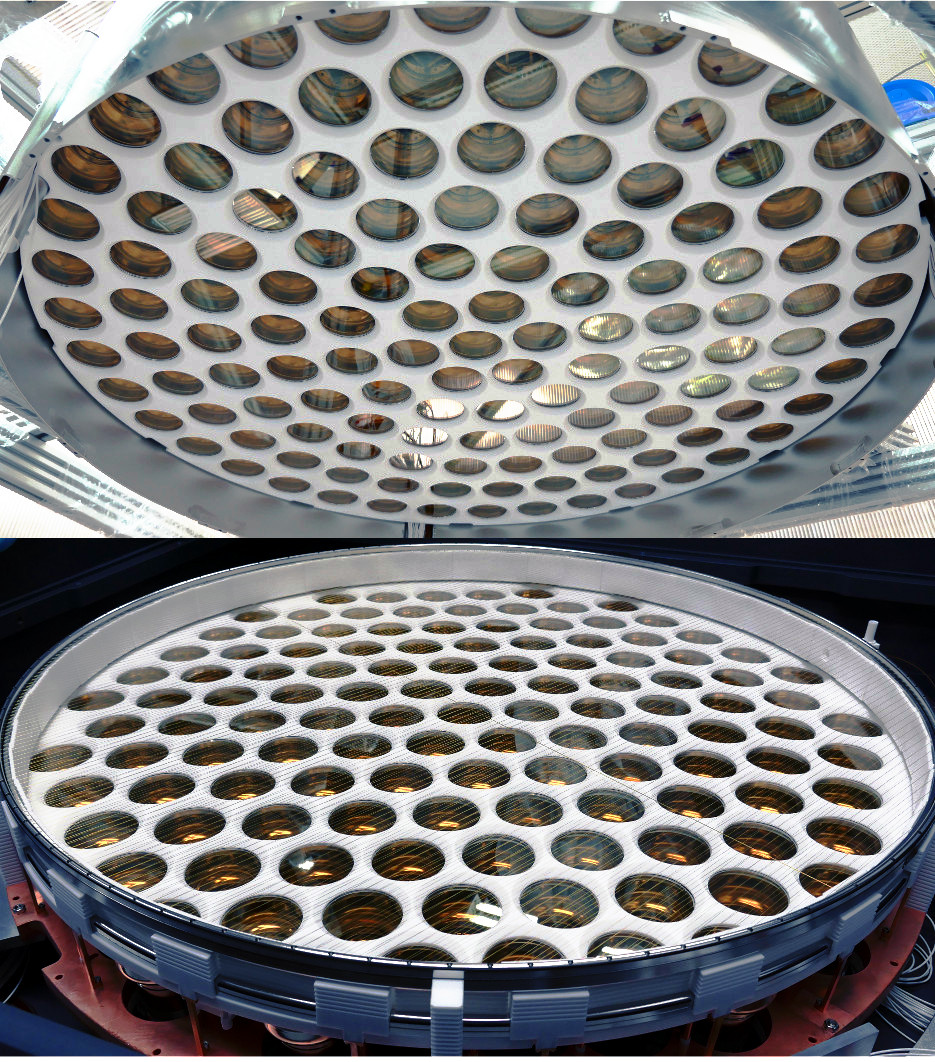}
\caption{The top and bottom PMT arrays are equipped with 127 and 121~Hamamatsu R11410-21 tubes, respectively. The top array is installed inside a diving bell, used to maintain and precisely control the LXe level. The bottom image also shows the cathode and the bottom screening electrode, installed in front of the PMTs. \label{fig::pmts}}
\end{figure}

\begin{figure*}[t!]
\centering
\includegraphics*[width=0.35\textwidth]{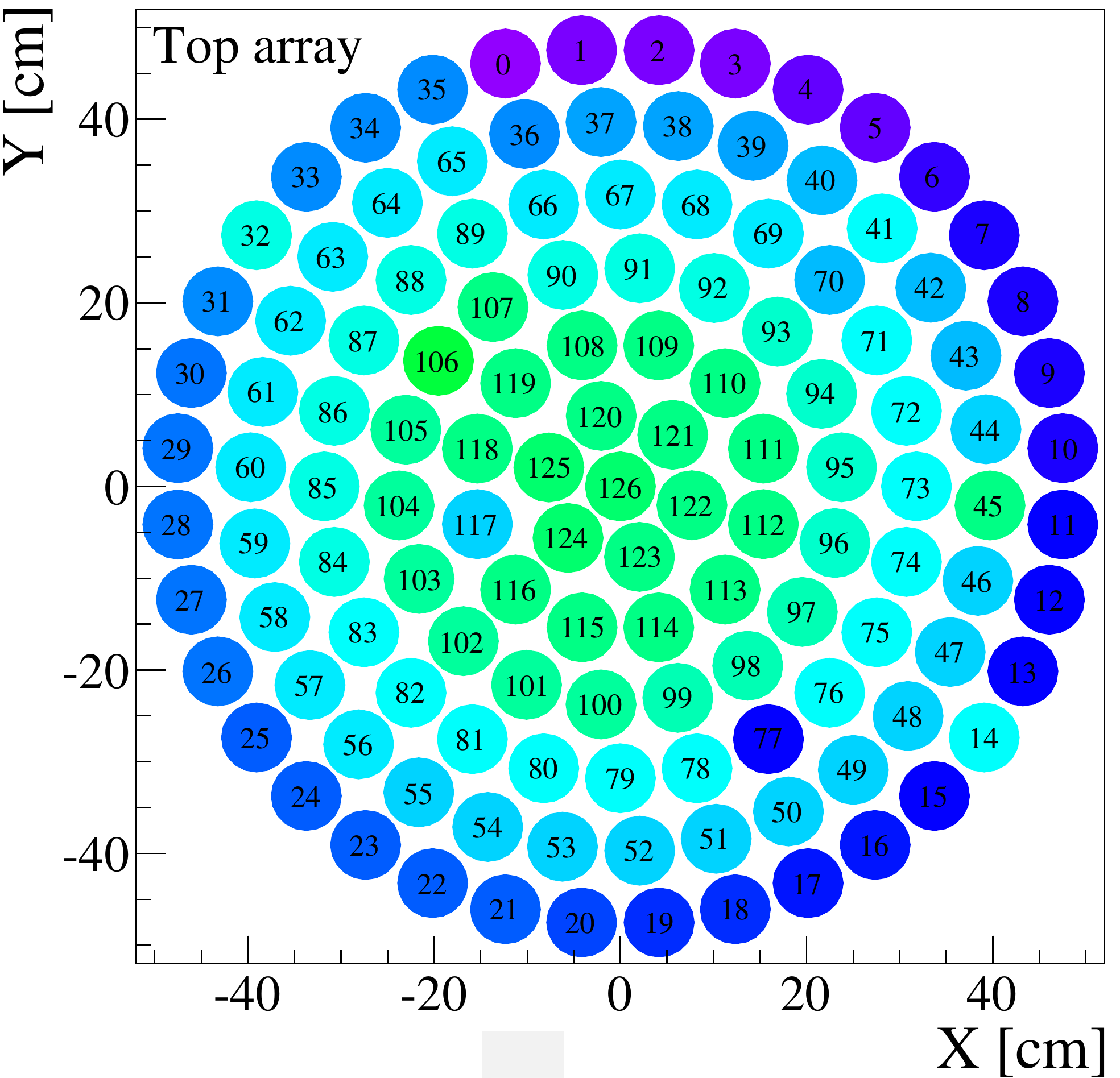} 
\includegraphics*[width=0.42\textwidth]{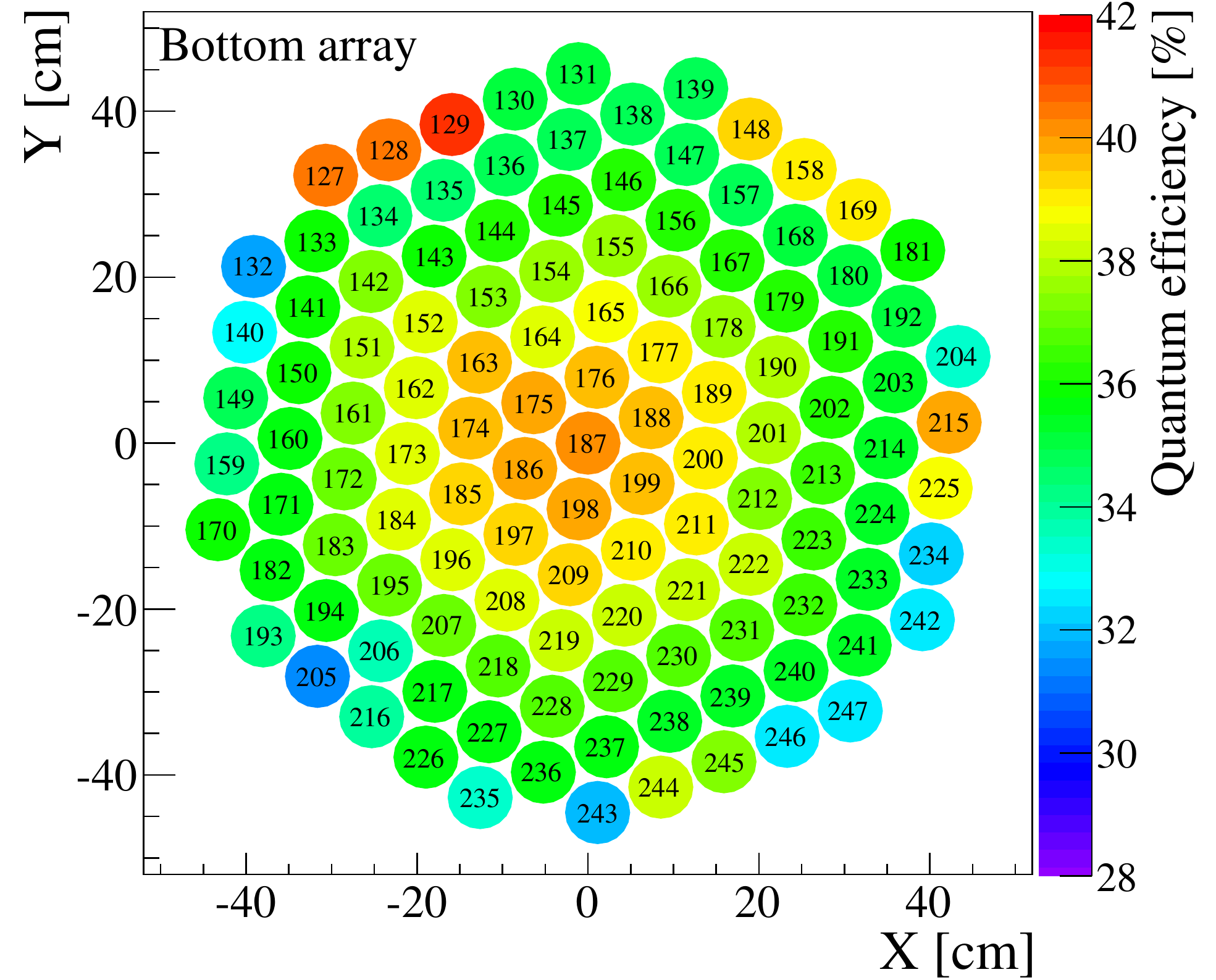} 
\caption{The PMTs were distributed in the top and bottom arrays mainly according to their quantum efficiency to ensure that the best-performing PMTs are located at the center of the bottom array, where most of the scintillation light is collected. Deviations take into account aspects such as increased radioactivity of individual tubes.\label{fig::qe}}
\end{figure*}

A ``diving bell'' made of stainless steel, which is directly pressurized by a controlled gas flow, is used to maintain a stable liquid-gas interface between the gate and anode electrodes. It encloses the top PMT array. The height of the liquid level inside the bell is controlled via a vertically-adjustable gas-exhaust tube. Possible tilts of the TPC are measured by means of four custom-made parallel-plate-capacitive levelmeters installed inside the diving bell. They cover a dynamic range of 10\,mm and have a precision of $\sim$30\,$\mu$m. Two cylindrical levelmeters of 1360\,mm length measure the LXe level during filling and recovery from below the cathode to above the bell, with 2\,mm precision.

%%%%%%%%%%%%%%%%%%%%%%%%%%%%%%%%%%%%%%%%%%%%%%%%%%%%%%%%%%%%%%%%%%%%%%
\subsubsection{Photomultipliers}
\label{sec::pmts}

A total of 248~Hamamatsu R11410-21 PMTs of 76.2\,mm~diameter are used to record the signals from the TPC. They are radially installed in the top array (127\,PMTs) to facilitate radial position reconstruction, and packed as tightly as possible in the bottom array (121\,PMTs) to maximize scintillation-light collection efficiency, see figure~\ref{fig::pmts}. They feature an average room-temperature quantum efficiency of 34.5\% at 178\,nm (bialkali-LT photocathode)~\cite{ref::r11410_character}, a high photoelectron collection efficiency of 90\% and are designed to operate stably in gaseous and liquid xenon at cryogenic temperature~\cite{ref::pmtpaper,ref::uclapmt}. The low-radioactivity PMT version~21 was jointly developed by Hamamatsu and the XENON collaboration~\cite{ref::r11410}. Six 25.4\,mm square-window Hamamatsu R8520 PMTs, as used in the XENON100 detector~\cite{ref::xe100_instr}, were installed in the LXe region outside of the TPC, next to the upmost field-shaping electrodes, for detector calibration studies~\cite{ref::xe100_analysis}.

All installed~R11410-21 PMTs were screened for their intrinsic radioactivity levels in batches of typically 15~tubes \cite{ref::r11410} and tested at room temperature and under gaseous nitrogen atmosphere at $-$100$^\circ$C. All PMTs were subject to at least two cooling cycles prior to installation. A subset of 44~tubes was additionally tested in LXe (2-3~cooling cycles). The PMTs feature a peak-to-valley ratio of $\sim$3, a single photoelectron resolution of 27\% for gains above $3 \times 10^6$ and a transit time spread (TTS) of $(9.1 \pm 1.3)$\,ns~\cite{ref::r11410_character}. A total of 73~tubes were rejected after the tests: 8~because of vacuum loss (``leak''), 53~because of emission of small amounts of light and 12~because of unstable dark count rates~\cite{ref::r11410_character}.

Based on the measured performance and the specifications provided by the manufacturer, the PMTs were placed in the two arrays. The PMTs with the highest quantum efficiency were installed at the center of the bottom array to maximize the light collection efficiency, see figure~\ref{fig::qe}. Both arrays consist of a massive OFHC copper support plate with circular cut-outs for the PMTs. A single PTFE plate holds the individual PMTs and a PTFE reflector-plate covers the areas between the PMT windows (see figure~\cite{fig::pmts}. Both PTFE plates are attached to the copper support in a self-centering way to ensure that all PMTs move radially inward upon cool-down to LXe temperatures while the support plate, which is connected to the remaining TPC structure, stays in place.

The 12\,dynodes of the R11410-21 PMT are connected to a custom-made low-background high-voltage divider circuit on a Cirlex printed circuit board. It was optimized for linear operation within the entire dynamic range of the ADCs (see section~\ref{sec::daq}), covering the entire energy region of interest for XENON1T ($\lesssim$\,1.5\,MeV). The signals are read via 50\,$\Omega$ RG196~coaxial cables. The PMTs are individually biased using CAEN A1535N and A1536N units via Kapton single-wire UHV cables (Accuglass 26\,AWG, TYP28-30), with the return current being routed through dedicated lines (2~redundant lines for 24~channels). Custom-developed low-pass filters installed on each high-voltage and return line reduce the electronic noise to sub-dominant levels. The cables were routed through the cryogenic pipe connecting the cryostat to the service building (see section~\ref{sec::cryostat}). The cables were potted into Conflat flanges (RH seals), with certified leak rates below $1\times 10^{-8}$\,mbar\,l/s, to ensure that the coaxial shields, as well as the high-voltage returns, remain separated from each other and from the detector ground. Installation was eased by integrating two custom-made low-background connectors for each bunch of 24~coaxial or 26~high-voltage cables (24~PMTs plus 2~return lines)~\cite{ref::kessler,ref::sivers}. One connector was placed above the TPC and the second one in front of the Conflat flanges in the gaseous xenon.

%%%%%%%%%%%%%%%%%%%%%%%%%%%%%%%%%%%%%%%%%%%%%%%%%%%%%%%%%%%%%%%%%%%%%%
\subsubsection{Cryogenics}
\label{sec::cryostat}

This section describes the cryostat, which contains the TPC with the LXe target, and the cryogenic system for gas liquefaction and compensating for heat losses.

\paragraph{Cryostat and Support Frame}

\begin{figure}[b!]
\includegraphics*[width=0.48\textwidth]{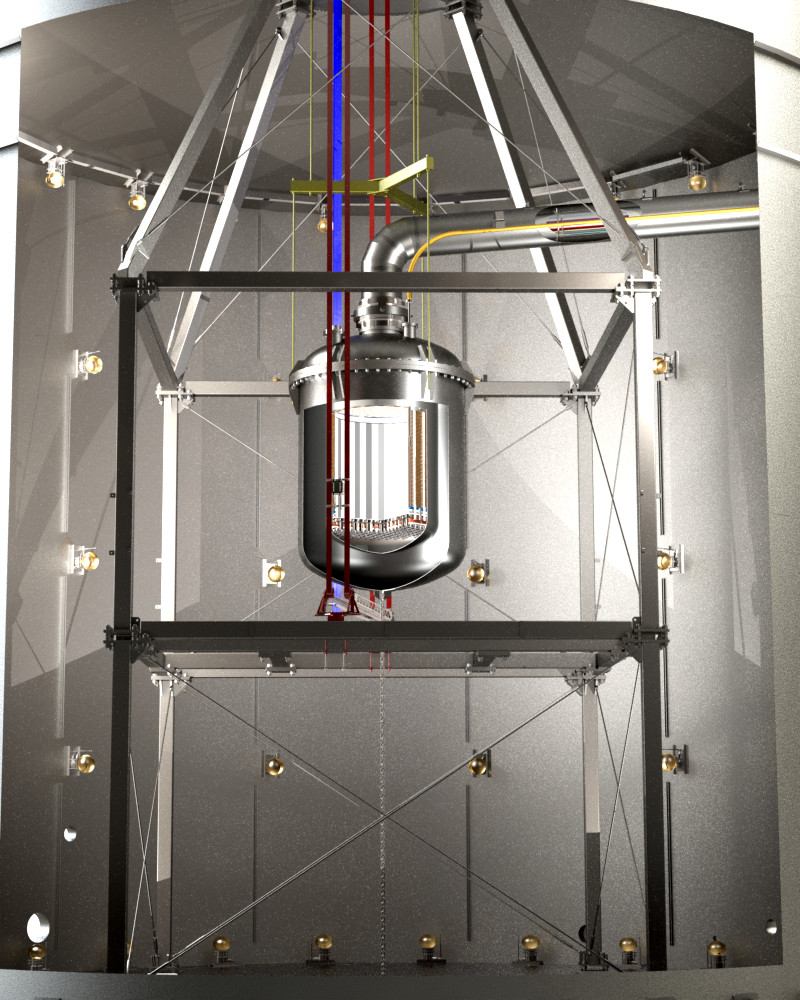}
\caption{The stainless-steel cryostat containing the LXe TPC is installed inside a 740\,m$^3$ water shield equipped with 84~PMTs deployed on the lateral walls. The cryostat is freely suspended (dark yellow) on a stainless-steel support frame, which can be converted into a cleanroom around the cryostat. The cryostat is connected to the outside by means of two pipes. The large, vacuum-insulated cryogenic pipe carries all gas/LXe pipes and cables. A small pipe (yellow) is used for the cathode high-voltage. Also shown is the system for calibrating XENON1T with external sources installed in movable collimators attached to belts (blue, red). \label{fig::wt}}
\end{figure}

The TPC is installed inside a double-walled, cylindrical stainless-steel cryostat made of low-radioactivity material~\cite{ref::screening}. The inner vessel is 1960\,mm high and 1100\,mm in diameter. Its inner surface, in direct contact with the liquid xenon, was electro-polished in order to reduce the emanation of radon. It is enclosed by an outer vessel of 2490\,mm height and 1620\,mm diameter, large enough to accomodate the detector of the upgrade stage XENONnT as well. The inner vessel is metal-sealed (Helicoflex) and thermally decoupled from the outer one by polyamid-imid (Torlon) spacers. Thirty~layers of superinsulation (RUAG Space Austria) reduce static thermal losses to $\sim$75\,W. The cryostat is installed in the center of the water Cherenkov detector (see figure~\ref{fig::wt} and section~\ref{sec::muonveto}). The connections to the outside are made through a double-walled cryogenic pipe (406\,mm~external diameter; 254\,mm inner diameter) enclosing all the connections to the cryogenic system (cooling, purification, fast emergency recovery, diving bell pressurization) and the cables for the PMTs and auxiliary sensors. A separate, single-walled pipe carries the high-voltage cable to the TPC cathode feedthrough.

The cryostat is freely suspended from three M20~threaded rods, attached to the top of a 10\,m high stainless-steel support frame erected inside the water tank. In order to minimize radioactive background from the frame, its components were selected for low radioactivity. The distance between the cryostat and the frame is at least 1\,m. The tilt of the cryostat, and thus the orientation of the TPC electrode planes with respect to the liquid xenon surface, can be adjusted from outside the water shield by changing the length of the three threaded rods. The precision of the tilt adjustment is better than 50\,$\mu$rad; the tilt is measured in real-time with a MEMS biaxial submersible tiltmeter (RST instruments). A chain connecting the cryostat to the water shield floor compensates buoyancy forces if the cryostat is empty.

A temporary 4.5\,$\times$\,4.5\,m$^2$ platform can be installed on the detector support frame at 3.2\,m height, to facilitate work on the detector. 

\begin{figure*}[t!]
\centering
\includegraphics*[width=1.0\textwidth]{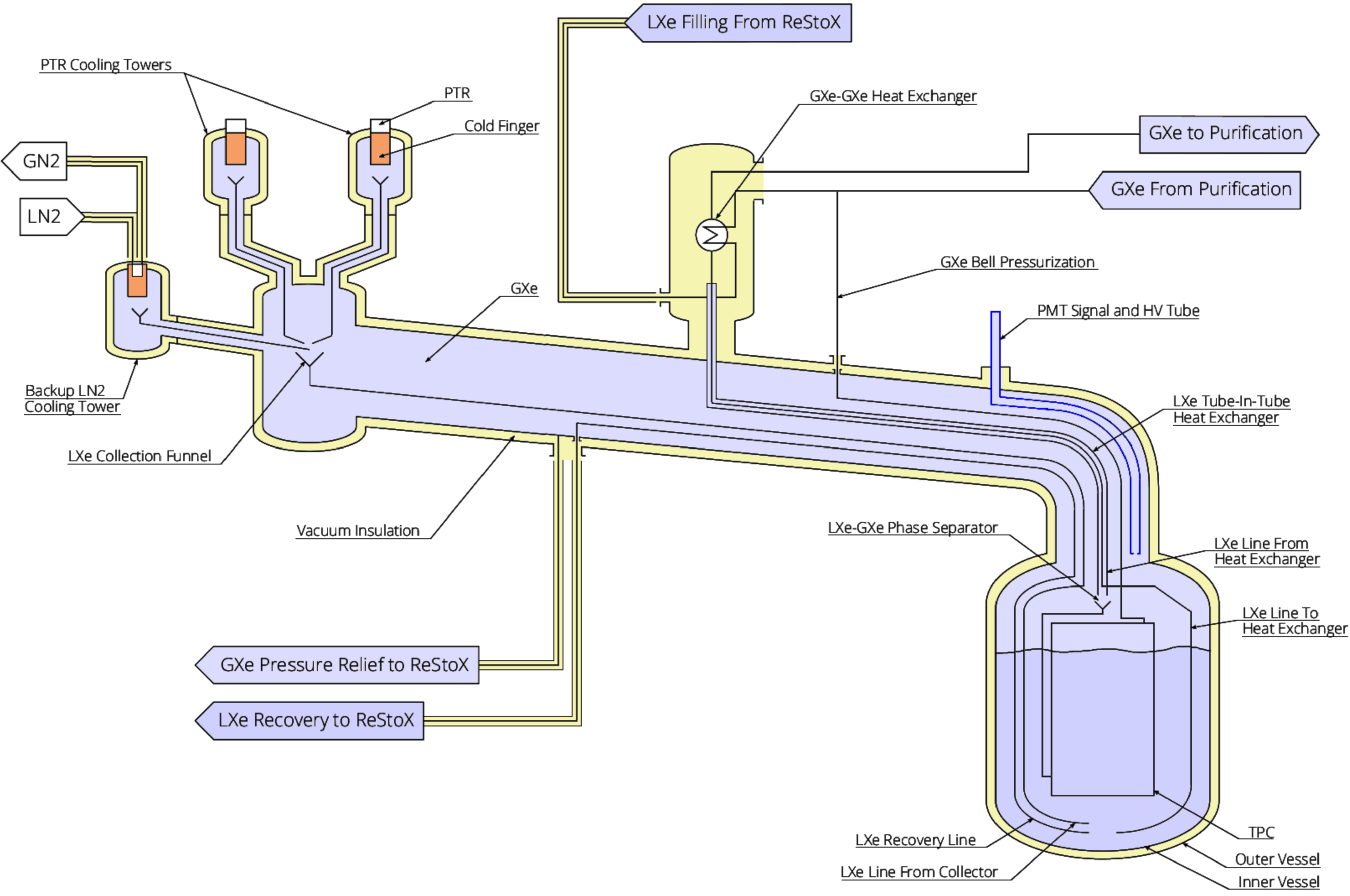}  
\caption{The cryogenic system of XENON1T: cooling is provided by means of three redundant cold heads (two pulse-tube refrigerators (PTR), 1~LN$_2$), installed on individual cooling towers located outside of the water shield. The liquefied xenon runs back to the main cryostat in a 6\,m long vacuum-insulated cryogenic pipe, through which all connections to the TPC are made with the exception of the cathode bias voltage which is not shown in the figure. The connections to the systems for xenon purification and storage (ReStoX) are also shown. Figure not to scale.\label{fig::cryo}}
\end{figure*}

%%%%%%%%%%%%%%%%%%%%%%%%%%%%%%%%%%%%%%%%%%%%%%%%%%%%%%%%%%%%%%%%%%%%%%1
\paragraph{Cooling}
\label{sec::cryogenics}

XENON1T follows the ``remote cooling'' concept that was successfully employed by XENON100~\cite{ref::xe100_instr}. It allows for maintenance of the cryogenic system, which is installed far away from the TPC, while the detector is cold. The xenon gas inside the XENON1T cryostat is liquefied and kept at its operating temperature $T_0 = -96^\circ$C by means of two redundant pulse-tube refrigerators (PTRs~\cite{ref::ptr}, Iwatani PC-150), which each provide $\sim$250\,W of cooling power at $T_0$. Each PTR is connected to a copper~cold finger reaching into the xenon~volume such that the PTR can be removed without exposing the inner vessel. The PTR insulation volumes are separated such that one PTR can be serviced while the other is in operation. The measured total heat load of the system is 150\,W, hence one PTR is sufficient to operate the detector. The xenon pressure inside the cryostat is kept constant by controlling the temperature of the active PTR cold finger using resistive heaters. A proportional-integral-derivative (PID) controller (Lakeshore~340) reads the temperature at the cold finger and controls the power supplied to the heaters. 

In case of a sudden pressure increase beyond a defined set point due to, e.g., a power loss, a PTR failure, or a partial loss of insulation vacuum, an additional backup liquid nitrogen (LN$_2$) cooling system maintains the pressure at a safe level. Its cold finger is cooled with a LN$_2$ flow and the cooling power is controlled by adjusting the nitrogen evaporation rate. The LN$_2$ is supplied by 
the same 10\,m$^3$ tank as used by the xenon storage system ReStoX~(see section~\ref{sec::xenon}). Only $\sim$100\,l/d are required to provide sufficient cooling power for XENON1T without PTRs. In normal operation, the backup LN$_2$ cooling system cold finger is kept a few degrees above the xenon liquefaction temperature. To ensure operation during a prolonged power loss, all safety-critical sensors and controllers for the emergency cooling system are powered by a uninterruptible power supply. 

The cryogenic system interfaces with the cryostat through the vacuum-insulated cryogenic pipe. Xenon gas from the inner cryostat vessel streams to the cryogenic system, is liquefied by the PTR, collected in a funnel and flows back to the cryostat vessel, driven by gravity, in a pipe that runs inside the cryogenic tube. Another pipe carries LXe out of the cryostat, evaporates it in a heat exchanger, and feeds it to the xenon purification system (see section~\ref{sec::xenon}). The purified xenon gas is liquefied in the same heat exchanger and flows back to the cryostat. The pipe that carries the purified LXe back to the cryostat is also used during the cryostat filling operation. Figure~\ref{fig::cryo} shows a schematic of the different components of the XENON1T cryogenic system and its interfaces to other systems.

%%%%%%%%%%%%%%%%%%%%%%%%%%%%%%%%%%%%%%%%%%%%%%%%%%%%%%%%%%%%%%%%%%%%%%
\subsubsection{Xenon Purification and Storage}
\label{sec::xenon}

\begin{figure*}[t!]
\includegraphics*[width=0.98\textwidth]{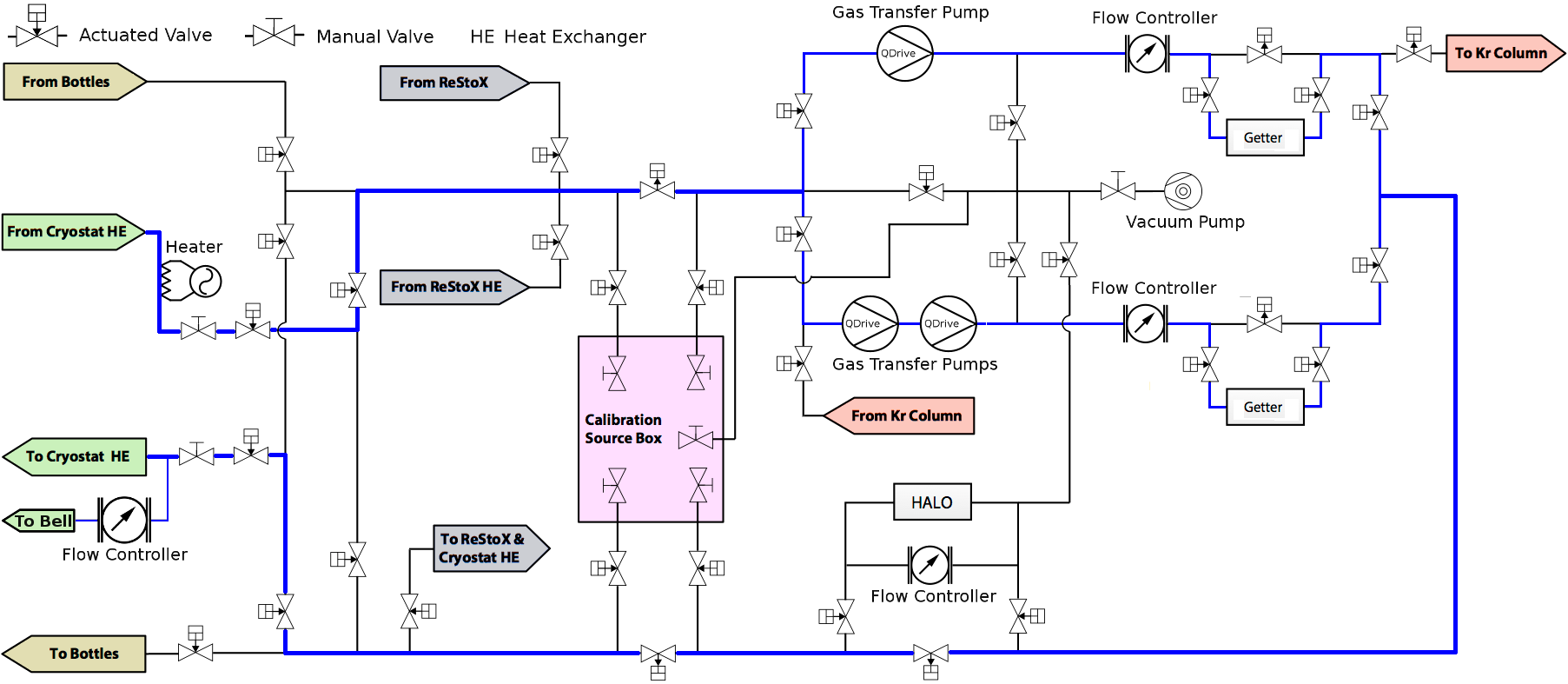}
\caption{Piping and instrumentation diagram (P\&ID) of the XENON1T purification system. The system also serves as the main interface to the other components of the gas-handling system (see figure~\ref{fig::cryosystem}) and allows the insertion of short-lived isotopes for calibration. Some instrumentation such as temperature and pressure sensors, as well as several access ports are omitted for clarity. The path of the xenon gas during standard purification is indicated in blue. \label{fig::pur}}
\end{figure*}

\begin{figure*}[t!]
\centering
\includegraphics*[width=0.65\textwidth]{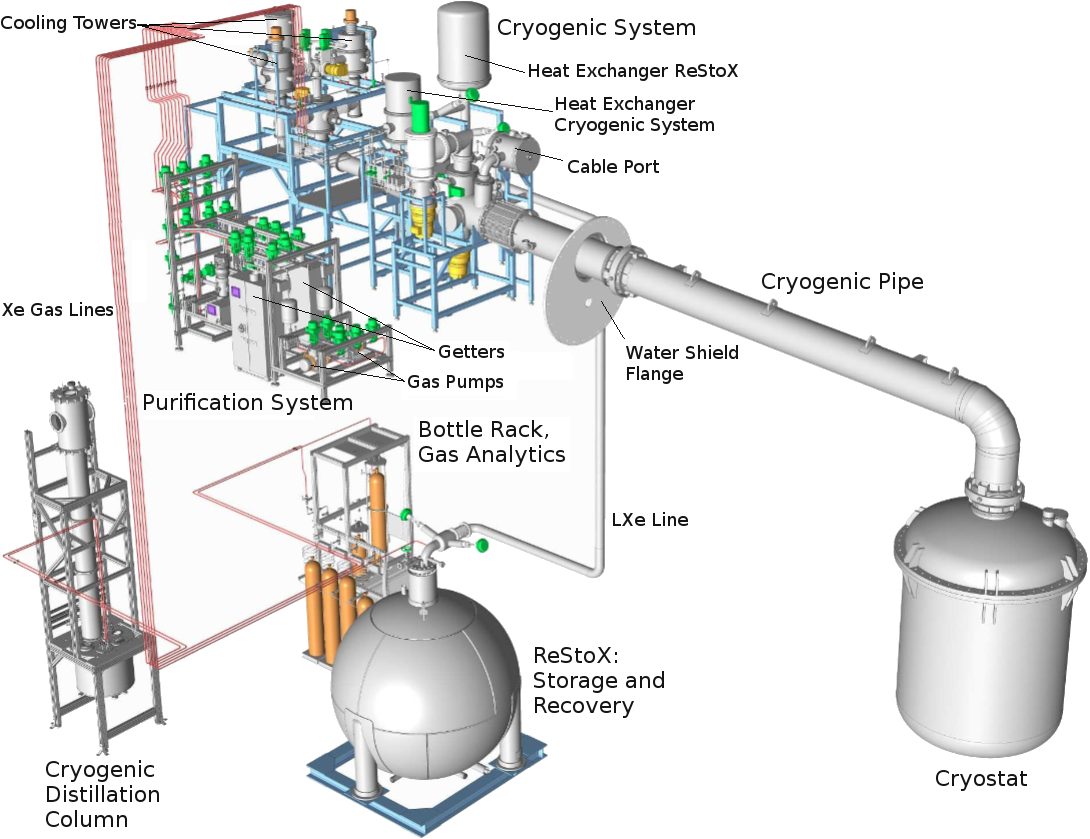}
\caption{The gas-handling system of XENON1T consists of the cryogenic system (cooling), the purification system (online removal of electronegative impurities), the cryogenic distillation column ($^\textnormal{\tiny nat}$Kr removal), ReStoX (LXe storage, filling and recovery), the gas bottle rack (injection of gas into the system) and gas analytics station (gas chromatograph). The cryostat inside the water shield accomodates the TPC. \label{fig::cryosystem}}
\end{figure*}

While the TPC encloses a LXe target of 2.0\,t, a total of 3.2\,t is required to operate the detector. The additional xenon lead is contained in the 60\,mm layer between the cathode electrode and the bottom PMTs, in the 58\,mm layer between the TPC field-shaping electrodes and the cryostat wall, in between and below the bottom PMTs, in the gas phase and in the gas purification and storage systems described below. 

The total xenon inventory from various suppliers comprises research-grade gas with a low concentration of impurities, especially $^\textnormal{\scriptsize nat}$Kr. The impurity level was measured in sets of four~gas bottles by gas chromatography (custom-optimized Trace GC Ultra from Thermo Fisher)~\cite{ref::pizzella}. In case a higher level than specified was detected, the bottles were purified using the distillation system (section~\ref{sec::kr}) before adding the gas to the storage system.

%%%%%%%%%%%%%%%%%%%%%%%%%%%%%%%%%%%%%%%%%%%%%%%%%%%%%%%%%%%%%%%%%%%%%%%%%%%%%%%%%%%%
\paragraph{Xenon Purification}
\label{sec::pur}

Electronegative impurities, such as water or oxygen, absorb scintillation light and reduce the number of ionization electrons by capture in an electron drift-time dependent fashion. These impurities are constantly outgassing into the xenon from all detector components. Therefore, the gas must be continuously purified to reduce the impurities to the $10^{-9}$~O$_2$-equivalent level~(ppb). Driven by gas transfer pumps, LXe is extracted from the cryostat at its bottom, next to the LXe condensate inlet from the cryogenic system. The LXe is evaporated in a heat exchanger system, made from a concentric tube in combination with a plate heat exchanger, which also cools the return gas from the purification system~\cite{ref::demonstrator_pur}. It is 96\%~efficient and reduces the heat input into the cryostat to only 0.39\,W/slpm (standard liters per minute).

Two redundant and independently serviceable purification loops are operated in parallel, see figure~\ref{fig::pur}. The components of one loop can be serviced or replaced without stopping the xenon purification. Each loop consists of a gas transfer pump (CHART QDrive; one loop is equipped with two pumps to improve operational conditions and stability), a mass-flow controller (MKS~1579A) and a high-temperature rare-gas purifier (getter, SAES PS4-MT50-R); the latter removes oxide, carbide and nitride impurities by forming irreducible chemical bonds with the getter material (zirconium). The high-capacity magnetic-resonance QDrive pumps feature a hermetically sealed pump volume and transfer the gas by means of externally-driven pistons oscillating in a compression space. Since all pistons, motors and valves are unlubricated, the QDrive is well-suited for high-purity applications. As the re-condensed, purified LXe flows back directly into the TPC, at two opposite locations below the cathode electrode (see also figure~\ref{fig::cryo}), a low $^{222}$Rn emanation of purification system is crucial for a low ER background, see also section~\ref{sec::bg}. 

More than 30~actuated pneumatic valves, shown in figures~\ref{fig::pur} and~\ref{fig::cryosystem} (green), are controlled by the slow control system (section~\ref{sec::slowcontrol}). Besides state-changes of individual components, it allows for automated changes between different operation modes. For safety reasons, a few manual valves were added at selected locations. Various pressure, temperature and other sensors are used to monitor the key parameters and instruments of the system, which was constructed from electropolished components and can be baked to 80-120$^\circ$C. Oil-free vacuum pumps allow for the evacuation of either the whole system or of individual sections, to ease servicing.

The purification efficiency can be monitored by a Tiger Optics HALO$+$~H$_2$O monitor, which measures the water concentration in the xenon gas, and can be useful for detecting possible leaks. The purification system is also used to inject calibration sources into the detector, which are dissolved in the xenon gas (see section~\ref{sec::calibration}).

\paragraph{Xenon Storage}

In the past, LXe detectors were filled by liquefying xenon from the gas phase and emptied by evaporating the liquid target. This technique poses operational challenges for experiments at the multi-ton scale. Filling XENON1T starting with xenon gas at 15$^\circ$C would require $\sim$2~months using 250\,W of cooling power. In addition, a fast recovery of the LXe in case of an emergency would be impossible.

\begin{figure}[t!]
\includegraphics*[width=0.48\textwidth]{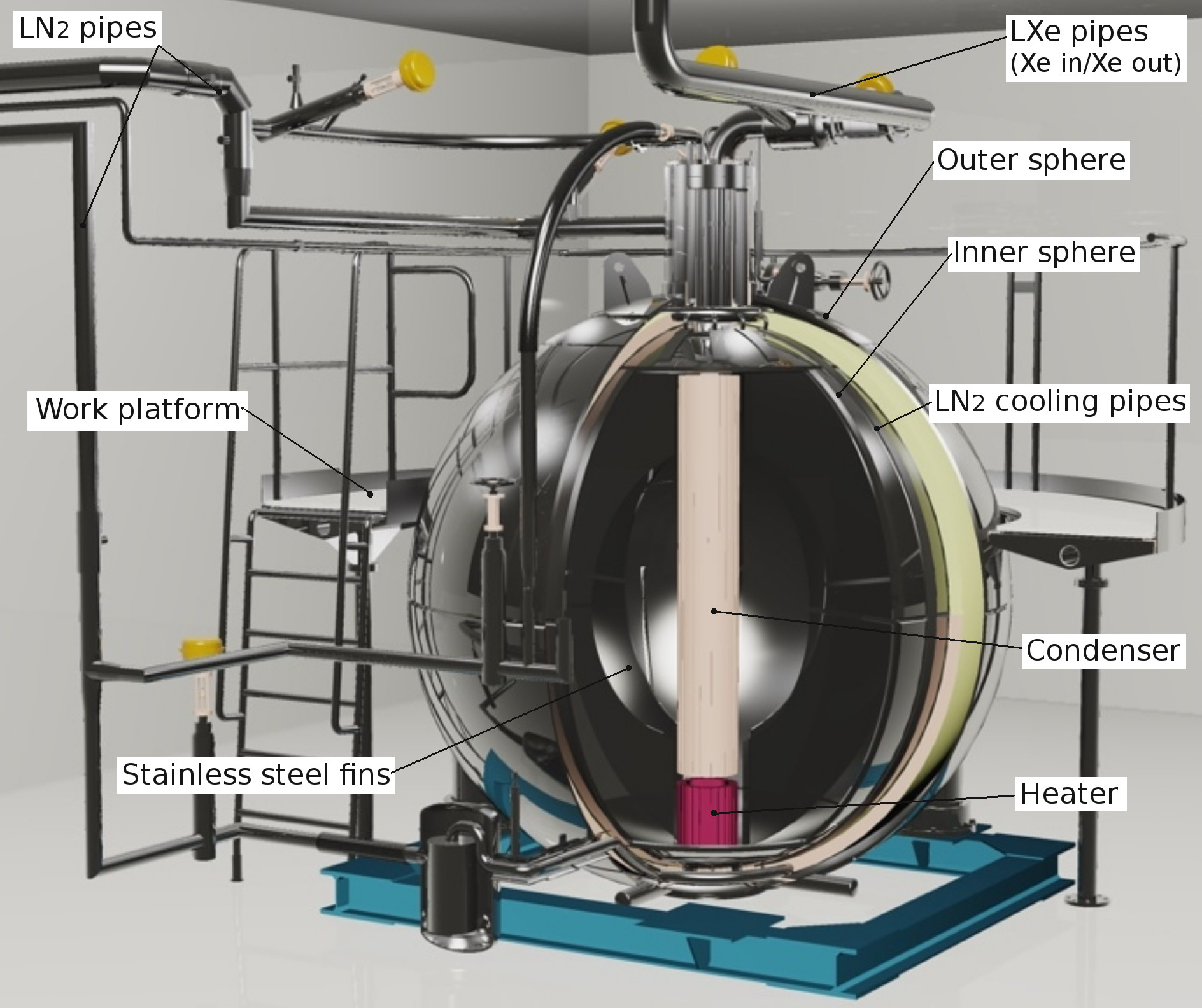}
\caption{Drawing of the xenon storage facility ReStoX, which can hold up to 7.6\,t of xenon in liquid or gaseous state. The condenser in the center of the 2.1\,m diameter sphere provides $>$3\,kW of cooling power. \label{fig::restox}}
\end{figure}

The newly developed xenon-storage system ReStoX~\cite{ref::restoxpatent} addresses these problems. It consists of a vacuum-insulated stainless-steel sphere with 2.1\,m diameter (4.95\,m$^3$ volume), see figure~\ref{fig::restox}. Its volume and the wall thickness of 28\,mm allow for storage of up to 7.6\,t of xenon as a liquid, as a gas and even as a super-critical fluid (being capable to withstand pressures up to 73\,bar). Superinsulation and minimized thermal contact between the inner and the outer cryostat spheres reduce the external heat load to 
$\sim$50\,W.

% Table: of main background sources and suppression mechanisms
\begin{table*}[t!]
\caption{Summary of the sources contributing to the background of XENON1T in a fiducial target of 1.0\,t and a NR energy region from 4 to 50\,keV (corresponding to 1~to 12\,keV ER equivalent). The expected rates are taken from the Monte Carlo simulation-based study~\cite{ref::xe1t_reach} and assume no ER rejection. CNNS stands for ``coherent neutrino nucleus scattering''.}
\label{tab::bgsources}
\centering
\small
\begin{tabular*}{\textwidth}{@{\extracolsep{\fill}}llll@{}}
\hline
Background Source & Type & Rate & Mitigation Approach \\ 
                  &      & [(t\,$\times$\,y)$^{-1}$] &  \\ \hline
$^{222}$Rn (10\,$\mu$Bq/kg) & ER & 620 & material selected for low Rn-emanation; ER rejection \\
solar pp- and $^7$Be-neutrinos & ER & ~~36 & ER rejection \\
$^{85}$Kr (0.2\,ppt of $^\textnormal{\tiny nat}$Kr) & ER & ~~31 & cryogenic distillation; ER rejection \\
2$\nu\beta\beta$ of $^{136}$Xe & ER & ~~~~9 & ER rejection \\
Material radioactivity & ER & ~~30 & material selection; ER and multiple scatter rejection; fiducialization \\ \hline
Radiogenic neutrons & NR & ~~~~0.55 & material selection; multiple scatter rejection; fiducialization \\
CNNS (mainly solar $^8$B-neutrinos) & NR & ~~~~0.6 & -- \\
Muon-induced neutrons & NR & ~$<$0.01 & active Cherenkov veto~\cite{ref::xe1t_mv}; multiple scatter rejection; fiducialization \\ \hline 
\end{tabular*}
\end{table*}

Cooling is achieved by means of LN$_2$, provided by an external 10\,m$^3$ dewar. A total of 16~LN$_2$ lines are welded to the outer surface of the inner vessel to cool down the sphere. Sixteen thin stainless-steel fins inside the volume additionally increase the heat exchange. In normal operation, i.e., while storing xenon in the liquid state, a condenser and heater system mounted in the center of the vessel precisely controls the pressure and ensures that the entrance pipe does not get blocked by  frozen xenon. Its cooling power of $>$3\,kW is distributed over a total of 4.3\,m$^2$ of copper surface.

The vessel and its cryogenic valves are all metal sealed and electropolished to allow for the storage of pre-purified LXe without sacrificing the target purity. To this end, ReStoX is connected to the detector (for filling and fast recovery) and to the purification system via an independent heat exchanger (for purification of the gas in ReStoX). The latter also provides access to the distillation column (for $^{85}$Kr removal, see section~\ref{sec::kr}). All components of the gas-handling system, their relative placement and connections are shown in figure~\ref{fig::cryosystem}. ReStoX is installed on the ground floor, about 7\,m below the top of the detector. The pumps of the purification system are used to transfer the xenon into the cryostat in a controlled way, at a speed of up to 50\,slpm: the LXe is evaporated in the ReStoX heat exchanger, purified, re-condensed in the same exchanger and transfered to the cryostat. The recovery of xenon into ReStoX via direct vacuum-insulated lines is driven by the pressure difference in the two systems. In case of emergency or for any recuperation of Xe gas, the detector pressure can be reduced within ${\cal O}$(1)\,minute.

%%%%%%%%%%%%%%%%%%%%%%%%%%%%%%%%%%%%%%%%%%%%%%%%%%%%%%%%%%%%%%%%%%%%%%
%%%%%%%%%%%%%%%%%%%%%%%%%%%%%%%%%%%%%%%%%%%%%%%%%%%%%%%%%%%%%%%%%%%%%%
%%%%%%%%%%%%%%%%%%%%%%%%%%%%%%%%%%%%%%%%%%%%%%%%%%%%%%%%%%%%%%%%%%%%%%
\subsection{Background Sources and Suppression}
\label{sec::bg}

The science goals of XENON1T require an unprecendented low background level~\cite{ref::xe1t_reach}. The main background sources are summarized in table~\ref{tab::bgsources}, divided into electronic (ER) and nuclear recoils (NR). The latter are most significant for the WIMP search, as single-scatter NR signatures from neutrons or neutrinos are indistinguishable from WIMP signals. 

Besides background suppression by shielding (see section~\ref{sec::muonveto}), material selection (section~\ref{sec::materials}) and active removal of radioactive isotopes (section~\ref{sec::kr}), backgrounds are effectively reduced in the data analysis: multiple scatter signatures are rejected 
based on the number of S2~peaks, ER-like events are identified based on the event's S2/S1~ratio, and external backgrounds, e.g., from radioactive decays in the detector construction materials or from muon-induced cosmogenic neutrons, are reduced by fiducialization, i.e., the selection of an inner detector region. However, fiducialization is not effective for target-intrinsic sources, such as the noble gases $^{222}$Rn and $^{85}$Kr, or the two-neutrino double-beta decay (2$\nu\beta\beta$) of $^{136}$Xe ($T_{1/2}=2.17 \times 10^{21}$\,y~\cite{ref::2nbb} with a 8.9\% abundance in $^\textnormal{\scriptsize nat}$Xe). It is also not effective for solar neutrino-induced backgrounds.

%%%%%%%%%%%%%%%%%%%%%%%%%%%%%%%%%%%%%%%%%%%%%%%%%%%%%%%%%%%%%%%%%%%%%%
\subsubsection{Water Shield and active Muon Veto}
\label{sec::muonveto}

An active water Cherenkov detector~\cite{ref::xe1t_mv} surrounds the cryostat. It identifies both muons, that have a flux of $(3.31\pm0.03)\times 10^{-8}$\,cm$^{-2}$s$^{-1}$ with an average energy of $\sim$270\,GeV in Hall~B of LNGS~\cite{ref::lvd}, and muon-induced neutrons by detecting 
showers originating from muon interactions outside the water shield. The water additionally provides  effective shielding against $\gamma$ rays and neutrons from natural radioactivity present in the experimental hall. The tank has a diameter of 9.6\,m and a height of 10.2\,m. The deionized water is provided by a purification plant (Osmoplanet DEMRO 2M~840), delivering up to 2.2\,m$^3$ of water per hour with a residual conductivity of 0.07\,$\mu$S/cm.

Operated as a Cherenkov muon veto, the water tank is instrumented with 84 PMTs of 20.3\,cm in diameter (Hamamatsu R5912ASSY) with a bialkali photocathode on a borosilicate window. The quantum efficiency is $\sim$30\% for wavelengths between 300~\,nm and 600\,nm, and the mean gain is $6\times10^6$ for a bias voltage of~1500\,V. The PMTs operate with a threshold that allows for the detection of single photoelectrons with $\sim$50\% efficiency. After optimization in a Monte Carlo study~\cite{ref::xe1t_mv}, the PMTs were deployed in five rings at the circumference of the water shield at different heights. The bottom ($z=0$\,m) and top ($z=10$\,m) rings consist of 24~evenly spaced PMTs, while only 12~PMTs are each installed in the three rings at $z=2.5$\,m, $z=5.0$\,m, and $z=7.5$\,m height. To further enhance the photon detection efficiency, the inner surface of the water tank was cladded with reflective foil (3M~DF2000MA) featuring a reflectivity of $>$99\% at wavelengths between 400\,nm and 1000\,nm~\cite{ref::response_MV}. The wavelength of the ultraviolet~Cherenkov photons can be shifted towards longer wavelengths in the reflection process to better match the PMT~sensitivity.

Each PMT can be calibrated by illumination with blue LED light through a plastic~fiber. In addition, the response of the full system can be measured by light emitted from four diffuser balls mounted on the cryostat support frame.

%%%%%%%%%%%%%%%%%%%%%%%%%%%%%%%%%%%%%%%%%%%%%%%%%%%%%%%%%%%%%%%%%%%%%%
\subsubsection[Detector Construction Materials]{Detector Construction Materials}
\label{sec::materials}

In order to reduce ER and NR background events, that arise from radioactive decays in the detector materials, all materials of the TPC, the cryostat and the support structure were selected for a low content of radioactive isotopes. Monte Carlo simulations were used to define the acceptable levels. The radioactivity measurements were performed using low-back\-ground high-purity germanium spectrometers of the XENON collaboration~\cite{ref::gempi,ref::gator,ref::giove}. The most sensitive spectrometers, located at the LNGS underground laboratory, can reach sensitivities down to the $\mu$Bq/kg level. In addition, standard analytical mass spectroscopy methods (ICP-MS, GD-MS) were employed at LNGS and at external companies. The measured radioactivity levels of the PMTs are summarized in~\cite{ref::r11410}; that of all other materials and components in~\cite{ref::screening}.

Most materials in contact with the liquid or gaseous xenon during standard operation were additionally selected for a low $^{222}$Rn emanation rate. This includes most components of the TPC, the inner cryostat and its connection pipes, the cryogenic system with its heat exchangers and the purification system. The LXe storage vessel and the cryogenic distillation column are irrelevant sources of Rn-emanation as they are not continuously connected to the TPC. Thus all $^{222}$Rn originating from these systems will rapidly decay to a negligible level. Even though the emanation rate is usually related to the $^{226}$Ra content of a material, which is obtained by $\gamma$ spectrometry, it must be measured independently since in most cases emanation is dominated by surface impurities. The measurements were performed according to the procedure described in~\cite{ref::rngerda} using the $^{222}$Rn emanation facility at MPIK Heidelberg~\cite{ref::rnemanation} and a similar one at LNGS. The results are summarized in~\cite{ref::masterrupp}.

To remove radioactive isotopes from surfaces, all TPC components were cleaned after production according to the following procedures: after degreasing, all copper pieces were pickled in a solution of 1\%~H$_2$SO$_4$ and 3\%~H$_2$O$_2$ and passivated in a 1\%~citric acid (C$_6$H$_8$O$_7$) solution. Afterwards the pieces were rinsed with de-ionized water and ethanol. The large stainless-steel pieces (diving bell, TPC electrode frames) were electropolished and cleaned with acetone, de-ionized water and ethanol. All small stainless-steel components (screws, rods, etc.) were degreased, pickled in a solution of both 20\%~HNO$_3$ and 2\%~HF, and finally passivated in a 15\%~HNO$_3$ solution before rinsing with de-ionized water and ethanol. The degreased PTFE components were immersed in a 5\%~HNO$_3$ solution and rinsed with de-ionized water and ethanol. Care was taken to not touch the reflecting TPC surfaces during cleaning, and all PTFE parts were stored under vacuum after the cleaning procedure. In cases of size limitations, the HNO$_3$-step was omitted and the sample was instead immersed in ethanol for a few hours.

The TPC was assembled above ground at LNGS, inside a custom-designed ISO\,5~class cleanroom with a measured particle concentration just above the ISO\,4~specification, using a movable installation and transport frame. The double-bagged TPC (aluminized mylar), fixed to the transportation frame, was moved to the underground laboratory by truck and attached to the top flange of the inner cryostat. A mobile class~ISO\,6 softwall cleanroom (4.5\,$\times$\,4.5\,m$^2$) was erected around the cryostat for this purpose.

%%%%%%%%%%%%%%%%%%%%%%%%%%%%%%%%%%%%%%%%%%%%%%%%%%%%%%%%%%%%%%%%%%%%%%
\subsubsection[Krypton Removal]{Krypton Removal by Cryogenic Distillation}
\label{sec::kr}

\begin{figure}[b!]
\centering
\includegraphics*[width=0.4\textwidth]{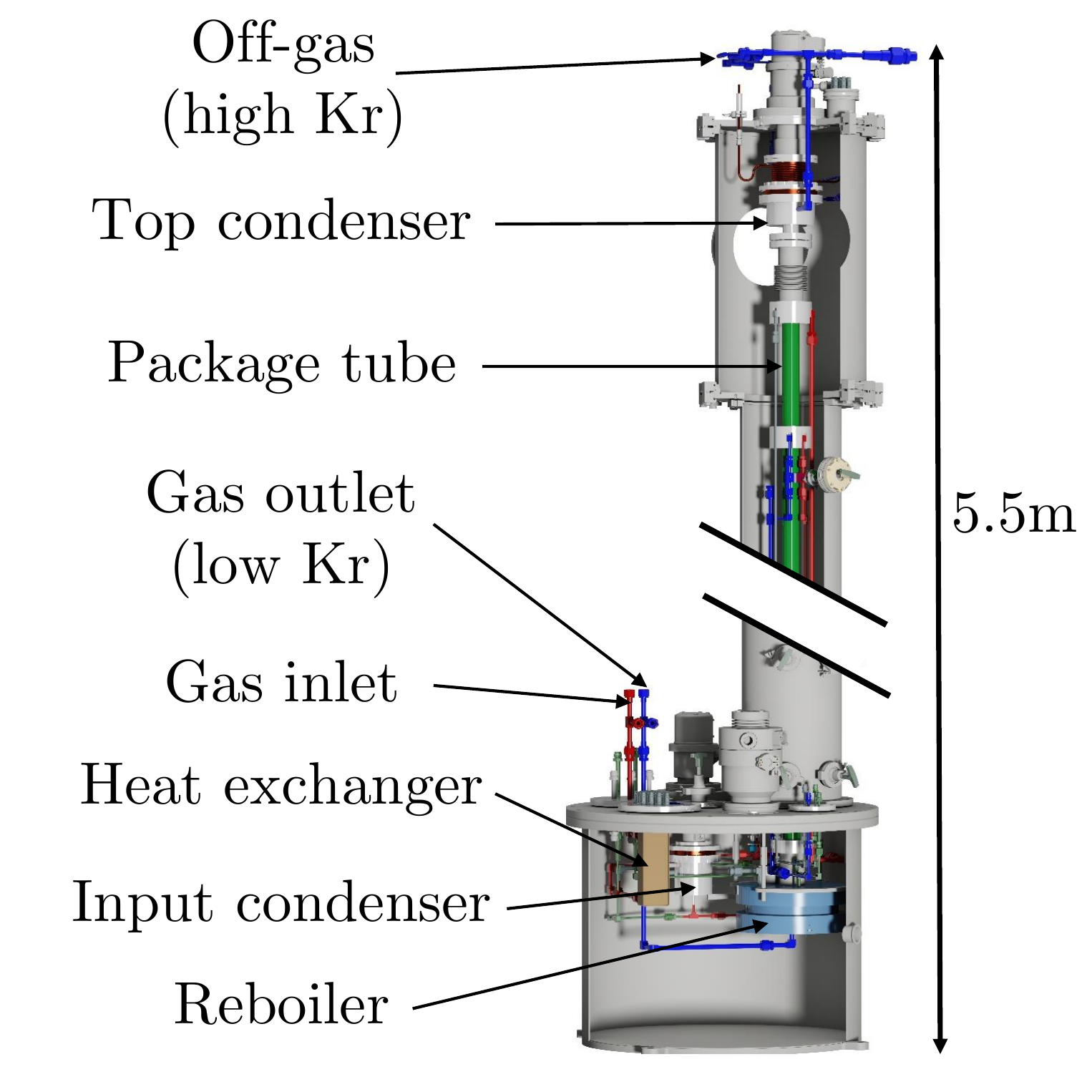}
\caption{The custom-designed XENON1T cryogenic distillation column. The outer vessels for vacuum insulation and most of the column package material were omitted for clarity. \label{fig::krcolumn}}
\end{figure}

Natural krypton, which contains the $\beta$-decaying isotope $^{85}$Kr ($T_{1/2}=10.76$\,y) at the $2 \times 10^{-11}$~level, is removed by cryogenic distillation, exploiting the 10.8~times larger vapor pressure of Kr compared to Xe at $-$96$^\circ$C. In a cryogenic distillation column, 
the more volatile Kr will hence be collected at the top while Kr-depleted Xe will be collected at the bottom. Given a $^\textrm{\scriptsize nat}$Kr/Xe concentration of $<$0.02\,ppm in commercial high-purity Xe gas, a Kr reduction factor around $10^5$ is required to reach the design goal of $^\textrm{\scriptsize nat}$Kr/Xe$<$0.2\,ppt. To achieve this goal, a distillation column using 2.8\,m of structured stainless-steel package material (Sulzer, type EX) was built following ultra-high vacuum standards. The total height of the XENON1T distillation system is 5.5\,m (see figure~\ref{fig::krcolumn}). The system is described in~\cite{ref::krcolumn} and can be operated stably at Xe gas flows up to 18\,slpm, corresponding to 6.5\,kg/h.

The Kr particle flux inside the column and the separation efficiency can be monitored using the short-lived isotope $^{83m}$Kr as a tracer~\cite{ref::tracer1,ref::tracer2}. After installation at LNGS, a separation factor of $(6.4^{+1.9}_{-1.4})\times 10^5$ was measured~\cite{ref::rga,ref::krcolumn}, reaching a concentration $^\textrm{\scriptsize nat}$Kr/Xe\,$<$\,0.026\,ppt and demonstrating that the system fulfills the requirements for XENON\-1T and for the future XENONnT. Such low concentrations are measured with a gas chromatography system coupled to a mass spectrometer (rare gas mass spectrometer, RGMS~\cite{ref::lindemann}). The possibility for online removal of~Rn was demonstrated first in a single stage setup~\cite{Bruenner:2016ziq} and by installing a shortened (1\,m package material) version of the final cryogenic distillation column in reverse and lossless mode on XENON100. A radon reduction factor of $>$27 (at 95\% CL) was achieved~\cite{ref::rnremoval}.

For the most efficient purification, the entire gas inventory would have to be transferred from ReStoX (section~\ref{sec::xenon}), via the distillation column, into the evacuated cryostat. This procedure would last for $\sim$3\,weeks, for a total Xe amount of 3.2\,t. However, to allow for data acquisition with a fully operational dual-phase TPC while at the same time reducing the Kr~concentration, the XENON1T collaboration has successfully established the online removal of~Kr. To this end, 7\%~of the purification gas flow was routed through the distillation column and the Kr-enriched gas (0.07\% of the total flow) was removed from the system. After continuously operating in this mode for 70\,days, with an initial measured $^\textnormal{\scriptsize nat}$Kr/Xe concentration of 60\,ppb, a final concentration of ($0.36\pm0.06$)\,ppt was measured by RGMS. This concentration is the lowest ever achieved in a LXe dark matter experiment. Being only a factor of~$\sim$2 above the XENON1T design goal, the concentration was sufficient for a first science run~\cite{ref::xe1t_sr0}.

%%%%%%%%%%%%%%%%%%%%%%%%%%%%%%%%%%%%%%%%%%%%%%%%%%%%%%%%%%%%%%%%%%%%%%
%%%%%%%%%%%%%%%%%%%%%%%%%%%%%%%%%%%%%%%%%%%%%%%%%%%%%%%%%%%%%%%%%%%%%%
%%%%%%%%%%%%%%%%%%%%%%%%%%%%%%%%%%%%%%%%%%%%%%%%%%%%%%%%%%%%%%%%%%%%%%
\subsection{TPC Calibration System}
\label{sec::calibration}

The PMT gains are calibrated by stimulating the emission of single photoelectrons from the photocathode by means of low-level light pulses from a blue LED. A total of four LEDs, installed in the counting room for easy accessibility, are simultaneously controlled by a 4-channel BNC-505 pulse generator. The light is guided into the cryostat via four optical fibers using SMA-SMA optical feedthroughs. Standard plastic fibers (980\,$\mu$m diameter core, light-tight jacket) are used externally. Bakeable synthetic silica fibers (600\,$\mu$m core, $-$190$^\circ$C to $+$350$^\circ$C) transfer the light to the cryostat. To reach uniform illumination of all PMTs and to minimize the calibration time, each of the silica fibers is split into six thin plastic fibers (250\,$\mu$m core) that feed the light into the TPC at various angular positions and heights around the field cage. A periodic external signal triggers the pulser and the TPC DAQ system; the LED calibration procedure is the only measurement which is not self-triggered (see also section~\ref{sec::daq}). 

Neutrons with energies around 2.2\,MeV and 2.7\,MeV from a Deuterium-Deuterium~(DD) fusion neutron generator (NSD Gradel Fusion NSD-35-DD-C-W-S) are used to calibrate the detector's NR response. By setting the generator voltage and current, the neutron flux can be tuned to the desired value. The generator was modified to achieve very low emission rates, around~10\,n/s in 4$\pi$ under stable conditions, as required for reducing the rate of pile-up events. The generator is stored outside of the water shield and can be displaced into it at three positions around the cryostat, to achieve a uniform illumination of the target. Details of the neutron generator system are given in~\cite{ref::n_gen}.

\begin{figure}[t!]
\includegraphics*[width=0.48\textwidth]{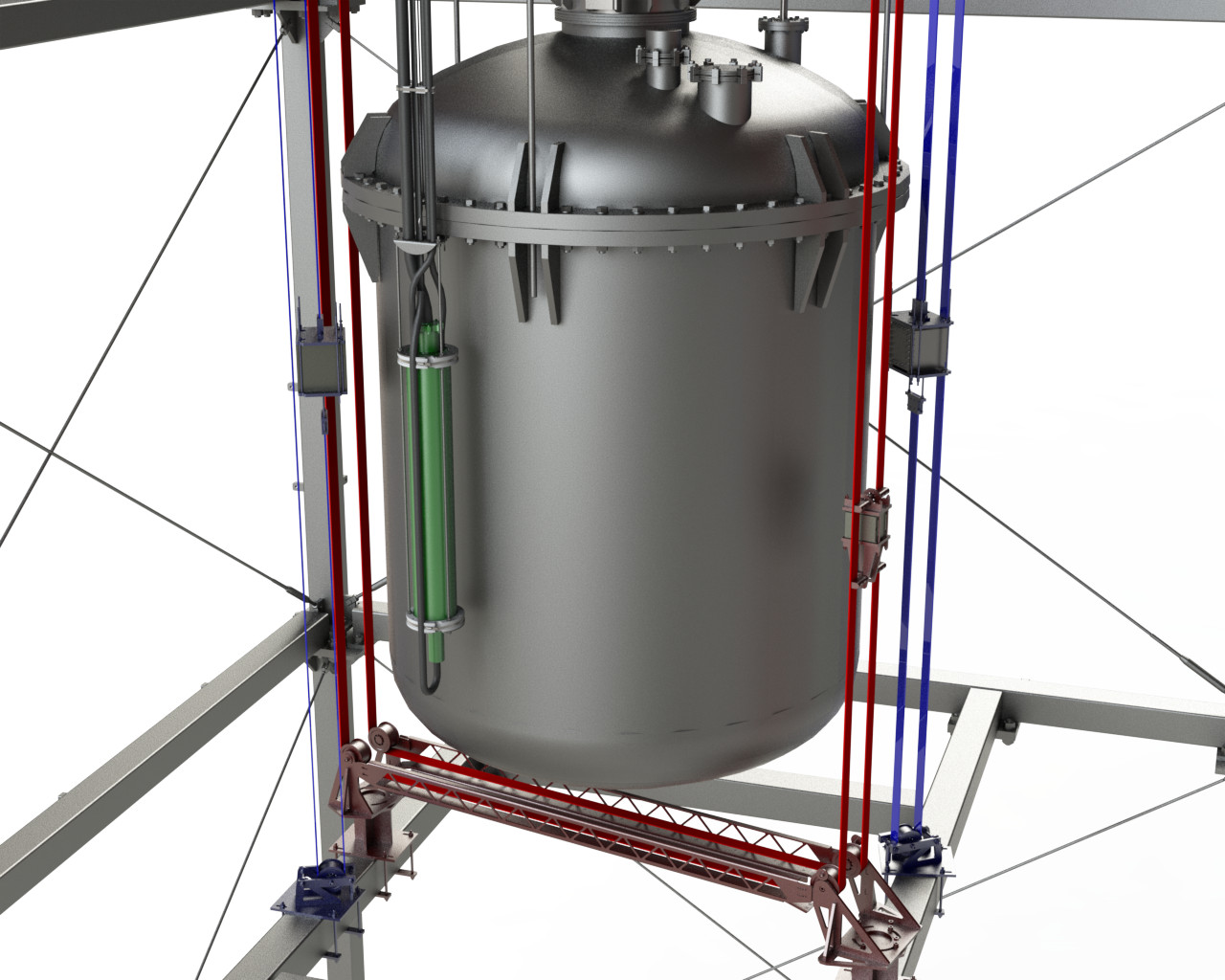}
\caption{The deployment system for external calibration sources. Two belts (``I-belts'', blue) allow for vertical movement of sources inside their W-collimators, while one belt (``U-belt'', red) reaches areas around the detector bottom. The DD-fusion neutron generator (green) can be vertically displaced along the cryostat. \label{fig::calibration}}
\end{figure}

$\gamma$ sources ($^{228}$Th, $^{137}$Cs) and a $^{241}$AmBe source to calibrate the ER and NR response, respectively, are installed in W-collimators; they can be deployed by means of belts from the top of the water shield to the cryostat. Two belts (``I-belt'', blue in figures~\ref{fig::wt} and~\ref{fig::calibration}) allow for moving the source vertically at two angular positions. Another belt (``U-belt'', red) crosses below the cryostat at $\sim$20\,cm distance from the central point. The collimators, which are stored above the water level when dark matter data are acquired, constrain the particles to a cone with 40$^\circ$-wide opening. This illuminates a central $\sim$1\,t fiducial volume when located at half height of the TPC. Residual $^{60}$Co activity in the steel of the cryostat flanges can also be used to assess the detector response to $\cal O$(1)\,MeV 
$\gamma$ rays. 

\begin{figure*}[t!]
\centering
\includegraphics*[width=0.90\textwidth]{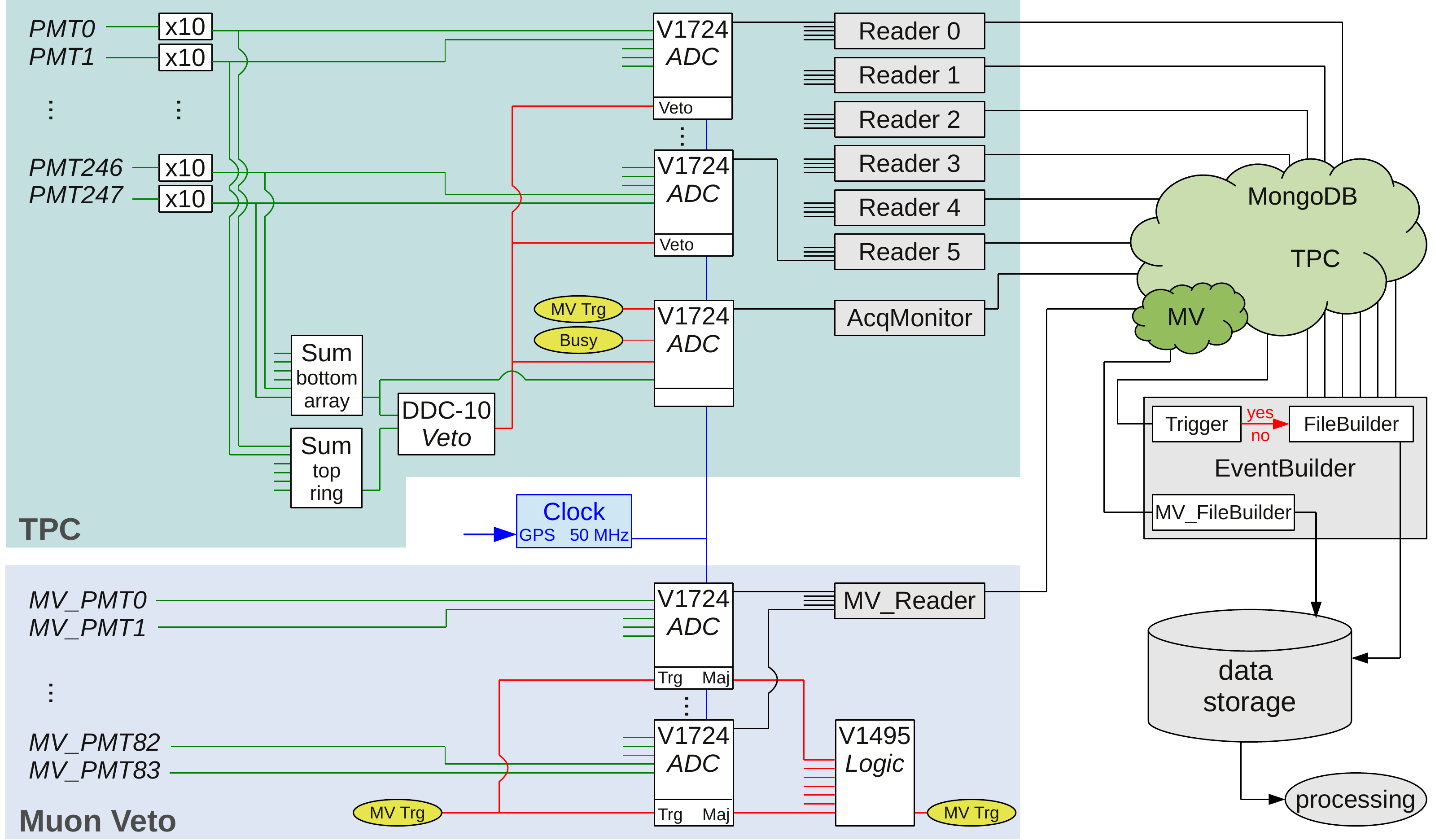}  
\caption{Illustration of the XENON1T DAQ system for the TPC and the muon veto. The two detectors are time-synchronized. The acquisition monitor consists of one high-data bandwidth ADC which records the sum of all bottom array PMTs and selected digital information, even when the other ADCs are busy or vetoed. \label{fig::daq}}
\end{figure*}

Due to the excellent self-shielding efficiency of LXe, the central target can only be calibrated with low-energy single-scatter ERs from dissolved sources. XENON1T uses $^{83m}$Kr ($T_{1/2}=1.8$\,h), the short-lived daughter of $^{83}$Rb, which delivers \monoenergetic \ conversion electron lines at 32.1\,keV and 9.4\,keV~\cite{Manalaysay:2009yq,Kastens:2009pa}. The $^{83}$Rb source is installed in the purification system to release the $^{83m}$Kr into the TPC when required. While the noble gas Kr mixes very well with Xe, it was shown that no long-lived $^{83}$Rb is emitted~\cite{Hannen:2011mr}. Tritiated methane (CH$_3$T), which delivers the tritium $\beta$ spectrum with the endpoint of 18.6\,keV and was pioneered by LUX~\cite{Akerib:2015wdi}, can also be injected into the XENON1T gas system. Due to its long half-life of 12.3\,y, it has to be removed from the LXe by the hot Zr-getters of the purification system~\cite{Dobi:2010ai}. The last intrinsic source is the noble gas isotope $^{220}$Rn ($T_{1/2}=56$\,s) which is efficiently emanated by an electrodeposited $^{228}$Th source ($T_{1/2}=1.9$\,y). The $^{220}$Rn decay chain produces $\alpha$, $\beta$ and $\gamma$ particles that are all useful for detector calibration~\cite{Lang:2016zde}, as demonstrated in XENON\-100~\cite{Aprile:2016pmc}. The $\beta$ decay of $^{212}$Pb (12.3\% branching ratio to the ground state, $Q=570$\,keV) delivers single-scatter ERs in the dark matter region of interest. Due to the rather short half-life $T_{1/2}=10.6$\,h of $^{212}$Pb, which dominates the chain, the activity is reduced by a factor $6 \times 10^{4}$~within one week.

%%%%%%%%%%%%%%%%%%%%%%%%%%%%%%%%%%%%%%%%%%%%%%%%%%%%%%%%%%%%%%%%%%%%%%
%%%%%%%%%%%%%%%%%%%%%%%%%%%%%%%%%%%%%%%%%%%%%%%%%%%%%%%%%%%%%%%%%%%%%%
%%%%%%%%%%%%%%%%%%%%%%%%%%%%%%%%%%%%%%%%%%%%%%%%%%%%%%%%%%%%%%%%%%%%%%
\subsection{Data Acquisition, Electronics and Computing}
\label{sec::daq}

The XENON1T TPC and the muon veto share a common data aquisition (DAQ) system; it can operate the two sub-detectors either simultaneously, during acquisition of dark matter search data, or separately for calibration. The overall DAQ scheme is illustrated in figure~\ref{fig::daq}. The PMT signals from the TPC and the muon veto are digitized by 100\,MHz CAEN~V1724 flash ADC boards with 14\,bit resolution, 40\,MHz bandwidth and a 2.25\,V or 0.5\,V dynamic range, respectively. The TPC channels are first amplified by a factor of~10 using Phillips Scientific~776 amplifiers (bandwidth: DC\,to\,275\,MHz). All ADCs share a common external clock signal to ensure that the two detectors and all digitizers are properly synchronized and share identical time stamps. The time signal can be optionally provided by a custom-developed module to obtain absolute GPS timing, relevant for the detection of supernova neutrinos~\cite{ref::shayne}. The module also provides a 0.1\,Hz synchronization signal. The DAQ is controlled via a web interface that also allows monitoring of the incoming data quality. Both DAQ systems are installed in the temperature stabilized XENON counting room and differ mainly in their trigger mode. 

The TPC DAQ is trigger-less in the sense that every pulse above a $\sim$0.3\,photoelectron (PE) digitization threshold, from every PMT, is read asynchronously and independently from all other channels. The baseline in-between such pulses is not digitized (zero suppression). To this end, a novel digitizer firmware was developed in cooperation with CAEN. Six computers (``readers'') are used for the parallel read-out of the 32~ADC boards, at a maximum rate of 300\,MB/s, corresponding to an event rate of $\sim$100\,Hz in calibration mode. The time-stamped digitized pulses are stored in a MongoDB noSQL database, along with some basic quantities of each pulse (time, channel). The sum signal of all bottom PMTs, generated by adding the individual signals by means of linear fan-in/fan-out modules, is continuously read by another computer (``acquisition monitor'') together with additional veto/busy information. The latter is used to precisely determine the measurement deadtime.

To reduce the input data rate during TPC calibration, a veto module based on a Skutek DDC-10 was developed. Depending on the size of the bottom array sum-signal or the relative amount of signal in the outer detector region, it issues a real-time veto signal which blocks the entire PMT data stream from being digitized. The digitizer firmware delays the incoming data for the required amount of time.

The trigger decision whether a particle interaction has occurred in the TPC is made in real-time by a software eventbuilder running on three server-grade machines (Fujitsu). It scans the MongoDB database for relevant signals, groups the data into causally connected events and stores them in a file. While a variety of trigger algorithms may be adapted to specific use cases, the standard dark matter and calibration S2~trigger is based on the time-clustering of pulses in individual PMT channels. A $>$99\% trigger efficiency is achieved at 200\,PE ($\approx$7\,e${^-}$). Meta-data on the trigger decision is stored with the raw data. It is available for online monitoring of the eventbuilder performance and offline analysis. 

The muon veto employs a classical coincidence trigger, managed by a custom-programmed CAEN V1495 VME unit, which requires at least $N_{\textnormal{\scriptsize pmt}}$ PMT signals in coincidence within a certain time window. The logic trigger signal is also sent to a channel of TPC acquisition monitor. For every muon veto PMT, the digitized waveform has a length of 5.12\,$\mu$s around the trigger signal. The data are written to the central MongoDB database and stored in files in the same way as the TPC data. 

Raw data from the DAQ system are temporarily moved to a buffer-storage at LNGS by an underground-to-aboveground connection using two 10\,Gbps fibers. Subsequently, the data are automatically transferred~\cite{ref::rucio} to dedicated storage points on the U.S.~Open Science Grid (OSG)~\cite{ref::osg} and the European Grid Infrastructure (EGI)~\cite{ref::egi}. The data are backed up in Stockholm. Data processing follows a similar philosophy and leverages the CI Connect service~\cite{ref::ci}. It allows for a unique submission portal while granting access to shared resources on OSG and EGI (using HTCondor~\cite{ref::htcondor} and glideinWMS services~\cite{ref::glideinWMS}) as well as dedicated allocations on local clusters at the member institutions. The data processor (see section~\ref{sec::pax}) derives high-level quantities from the digitized waveforms which are stored in files accessible through a JupyterHub infrastructure~\cite{ref::jupyter}. The data quality is constantly monitored by the DAQ system (noise, baselines, trigger, etc.). Certain high-level quantities such as electron lifetime or light yield are computed offline and monitored as well.

%%%%%%%%%%%%%%%%%%%%%%%%%%%%%%%%%%%%%%%%%%%%%%%%%%%%%%%%%%%%%%%%%%%%%%
%%%%%%%%%%%%%%%%%%%%%%%%%%%%%%%%%%%%%%%%%%%%%%%%%%%%%%%%%%%%%%%%%%%%%%
%%%%%%%%%%%%%%%%%%%%%%%%%%%%%%%%%%%%%%%%%%%%%%%%%%%%%%%%%%%%%%%%%%%%%%
\subsection{Slow Control System}
\label{sec::slowcontrol}

The various XENON1T subsystems and their instruments are operated, controlled and their status are monitored and recorded by a slow control system which is based on industry-standard process control hardware and software from General Electric (GE): Programmable Automation Controllers (PACs) for hardware and  Cimplicity SCADA (Supervisory Control And Data Acquisition) for software. Alarm conditions (e.g., parameter out of range, equipment failure, connection loss, etc.) are notified by email, cellular phone  SMS~\cite{ref::nexmo} and pre-recorded voice messages via a landline. The values of nearly 2500~parameters are stored in a GE Proficy Historian database, which offline analysis programs may query via a custom-developed Web API. The alarm notification, slow control viewer and offline analysis tool were custom-developed to complement the GE functionality.

The sensors and actuators of the cryogenics, LXe purication, LXe storage, Kr distillation, and water purification systems are controlled via individual PACs (GE RX3i family) that are connected to a private front-end network. Exceptions at PAC level are communicated to the alarm system using the GE Alarm\&Event Express tool. Local operation by means of touch screens is also possible should the SCADA system be unavailable. The high-voltage supplies, the DAQ system and the motor controllers of the calibration system are integrated into the slow control system via industry standard Open Platform Communication (OPC) servers, the Modbus protocol or web services. Potentially unsafe operations are additionally ``guarded'' by requiring specific conditions to be met before the operation can be executed.

Two redundant SCADA servers in active-passive fail-over mode connect to the PACs and OPC servers on the  private front-end network. All supervisory and data storage elements, such as the Historian database, the alarm system, the slow control viewer as well as the central XENON1T control room in an aboveground building are connected to the private back-end network. Two dedicated, redundant fiber links connect the experiment underground with the aboveground laboratory. In case of failure of the laboratory network, the slow control system is directly connected to a backup network at a point outside of LNGS. For safety reasons, the entire slow control system is powered by a dedicated uninterruptable power supply with extended on-battery runtime and generator backup. The system is protected by a firewall and only authorized users have the possibility to perform operations beyond data access, according to their pre-defined role. More details on the system are presented in~\cite{ref::sc}.

%%%%%%%%%%%%%%%%%%%%%%%%%%%%%%%%%%%%%%%%%%%%%%%%%%%%%%%%%%%%%%%%%%%%%%
%%%%%%%%%%%%%%%%%%%%%%%%%%%%%%%%%%%%%%%%%%%%%%%%%%%%%%%%%%%%%%%%%%%%%%
%%%%%%%%%%%%%%%%%%%%%%%%%%%%%%%%%%%%%%%%%%%%%%%%%%%%%%%%%%%%%%%%%%%%%%
\section{Detector Commissioning Results}
\label{sec::commissioning}

\begin{figure}[h!]
\includegraphics*[width=0.48\textwidth]{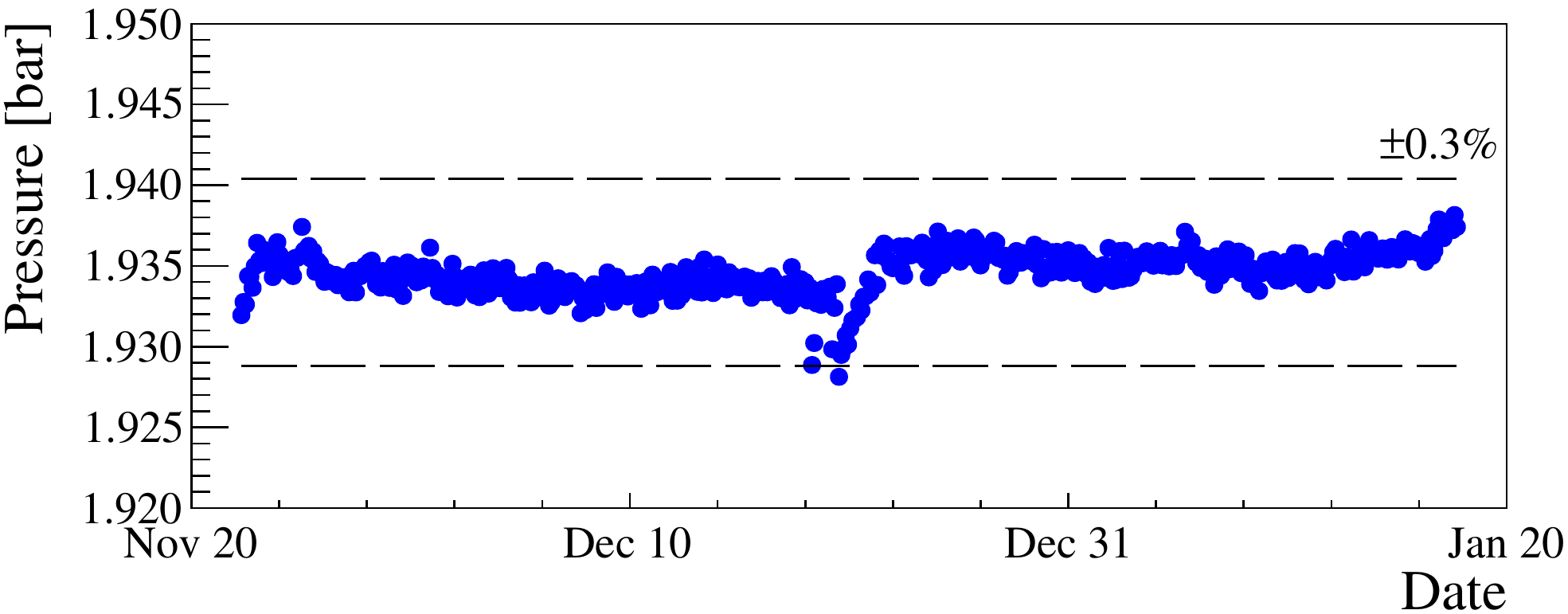}
\includegraphics*[width=0.48\textwidth]{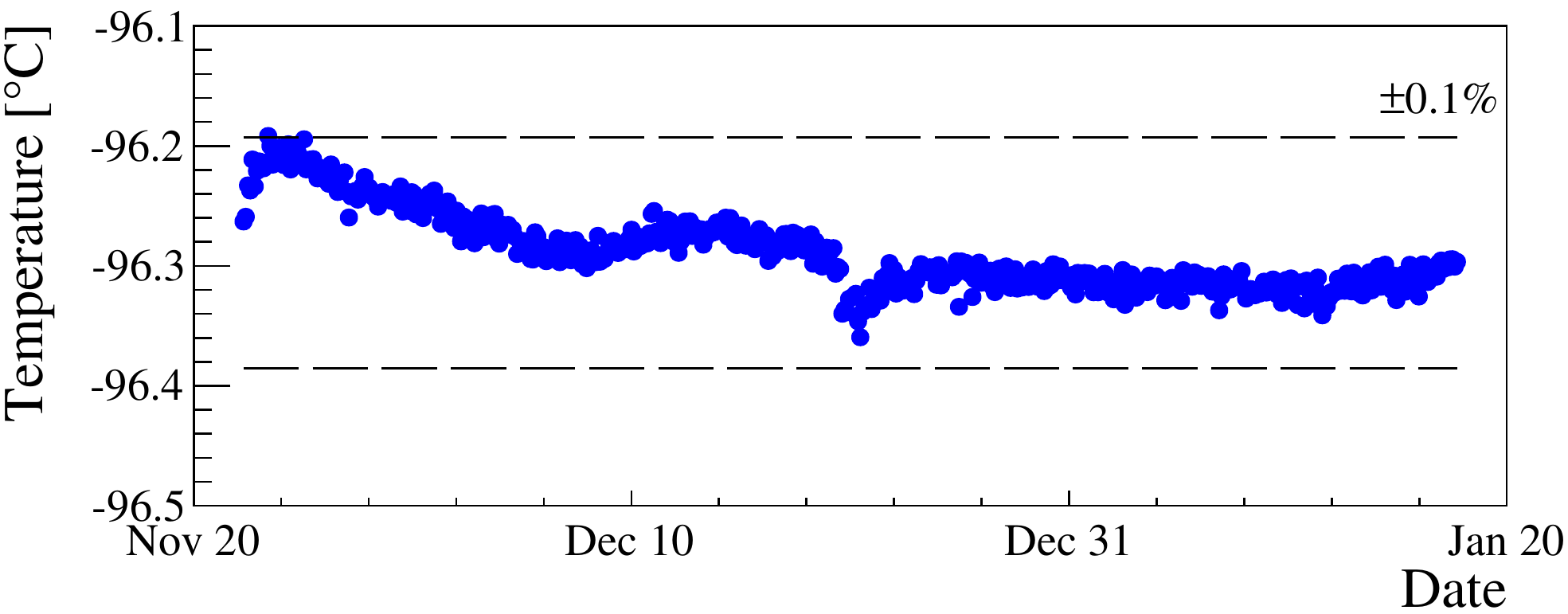}
\caption{Xenon gas pressure (top) and liquid xenon temperature at the TPC's half-height (bottom) measured by the XENON1T slow control system over a period of two months, during \srzero. The dip in temperature and pressure around December~19 is from a well-understood change in operating configuration, starting a Rn distillation campaign. \label{fig::sc_stability}}
\end{figure}

This section reports on the performance of the XENON1T detector during commissioning in summer 2016. It focuses especially on \srzero, which is comprised of 34.2~live days of data acquired between November 2016 and January 2017. The detector was operated under stable conditions over that period (see figure~\ref{fig::sc_stability}). The result of this run led to the most stringent exclusion of spin-independent WIMP-nucleon scattering interactions for WIMP masses $m_\chi>10$\,GeV/$c^2$, with a minimum of $7.7 \times 10^{-47}$\,cm$^2$ for $m_\chi=35$\,GeV/$c^2$ (90\% CL)~\cite{ref::xe1t_sr0}.

With a rate of $(70 \pm 9)$\,(t$\times$y$\times$keV)$^{-1}$ in the low-energy region of interest for WIMP searches and inside a 1.0\,t fiducial target, the ER background achieved in \srzero \ is the lowest ever reached in a dark matter experiment. It agrees with the prediction of $(84 \pm 7)$\,(t$\times$y$\times$keV)$^{-1}$, where the Monte Carlo result~\cite{ref::xe1t_reach} was updated for the measured Kr-concentration. This demonstrates that the XENON1T goal for the dominant background source, namely a $^{222}$Rn concentration around 10\,$\mu$Bq/kg (see table~\ref{tab::bgsources}), was also reached. Due to the short exposure, NR backgrounds were irrelevant for \srzero.

%%%%%%%%%%%%%%%%%%%%%%%%%%%%%%%%%%%%%%%%%%%%%%%%%%%%%%%%%%%%%%%%%%%%%%
\subsection{Muon Veto Performance}

The water shield of XENON1T was first filled in mid July 2016. The gains of the 84~PMTs in the water were equalized to $6.2 \times 10^6$. For \srzero, the muon veto trigger was slightly modified compared to the configuration presented in~\cite{ref::xe1t_mv}: the coincidence condition was increased from a PMT number of $N_\textnormal{\scriptsize pmt}=4$ to~$N_\textnormal{\scriptsize pmt}=8$ above a threshold of 1\,PE in a 300\,ns window. This increase has little impact on the muon tagging efficiency as the trigger rate of $R=0.04$\,Hz is constant above $N_\textnormal{\scriptsize pmt}=11$, indicating that nearly all crossing muons are detected. The measured rate corresponds to 144\,muons/h and agrees with the expectations. The trigger rate at $N_\textnormal{\scriptsize pmt}=8$ is 0.35\,Hz, significantly smaller than the $\sim$80\,Hz when triggering with $N_\textnormal{\scriptsize pmt}=4$, which reduces the total amount of data considerably. The rate increase toward lower coincidence levels is due to $\gamma$ rays from natural radioactivity, dominated by the primordial isotopes $^{238}$U, $^{232}$Th and $^{40}$K in the rock, 
concrete and structures of the Hall~B of LNGS. The observed $\gamma$ ray flux is 
in agreement with direct measurements performed at the same location~\cite{Haffke:2011fp}.

\begin{figure}[b!]
\includegraphics*[width=0.48\textwidth]{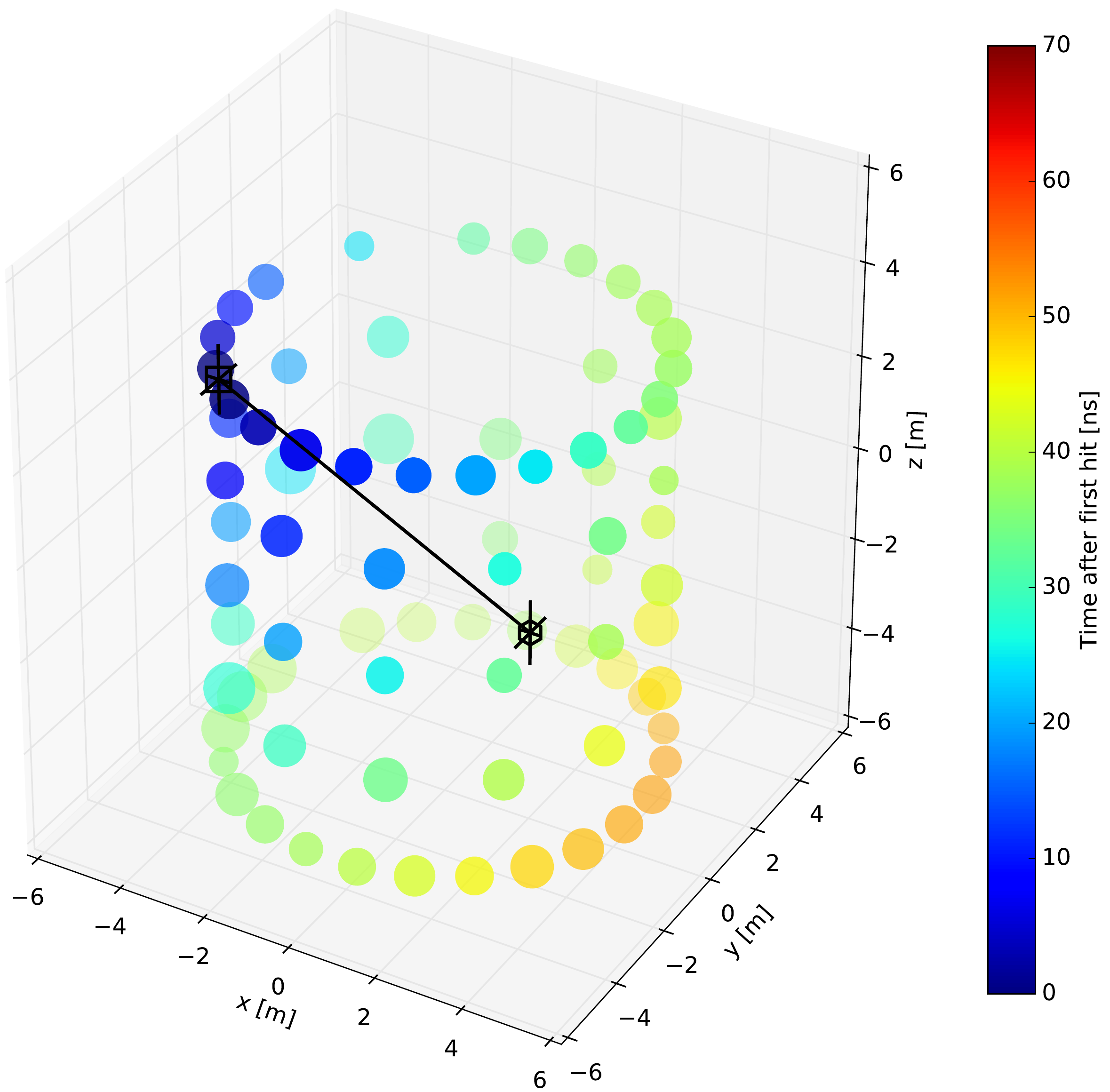}
\caption{The arrival time information of light in the PMTs of the muon veto detector (color coded points) allows the approximate reconstruction of the muon track (black line) though the water shield. This example shows an event where the muon traversed the shield close but next to the TPC. \label{fig::mvevent}}
\end{figure}

\begin{figure*}[p!]
\centering
\includegraphics*[width=0.77\textwidth]{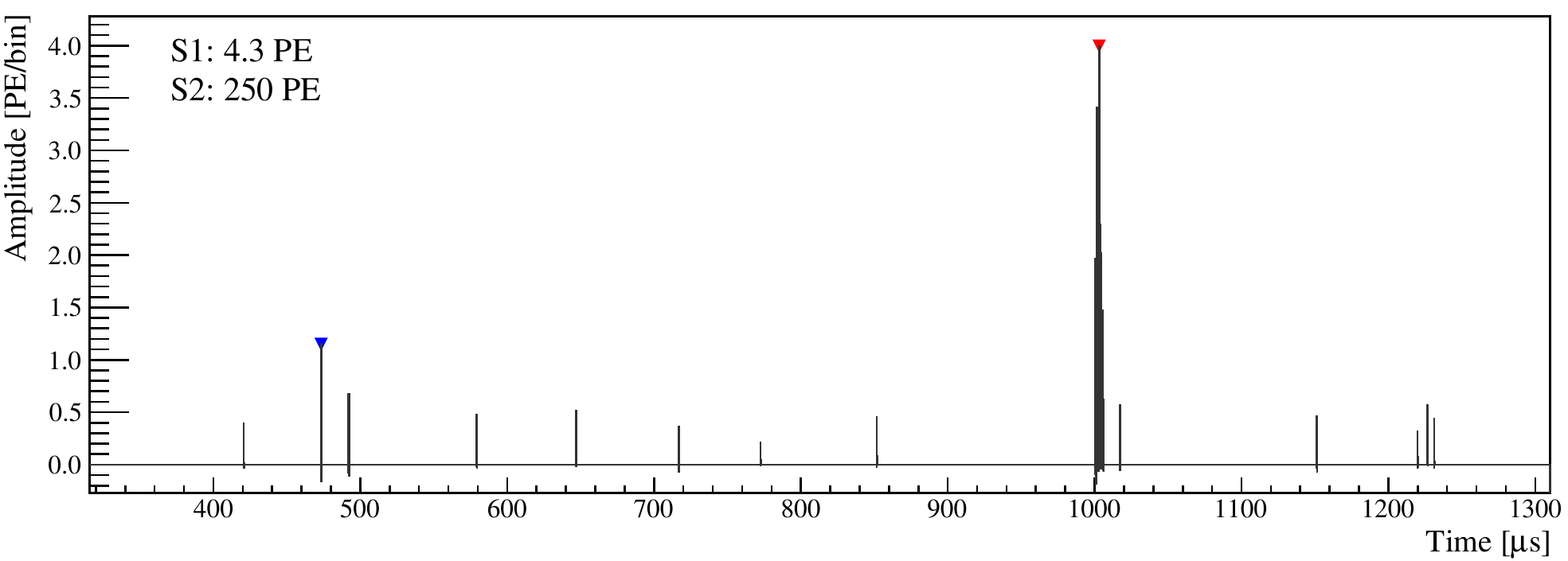}
\includegraphics*[width=0.38\textwidth]{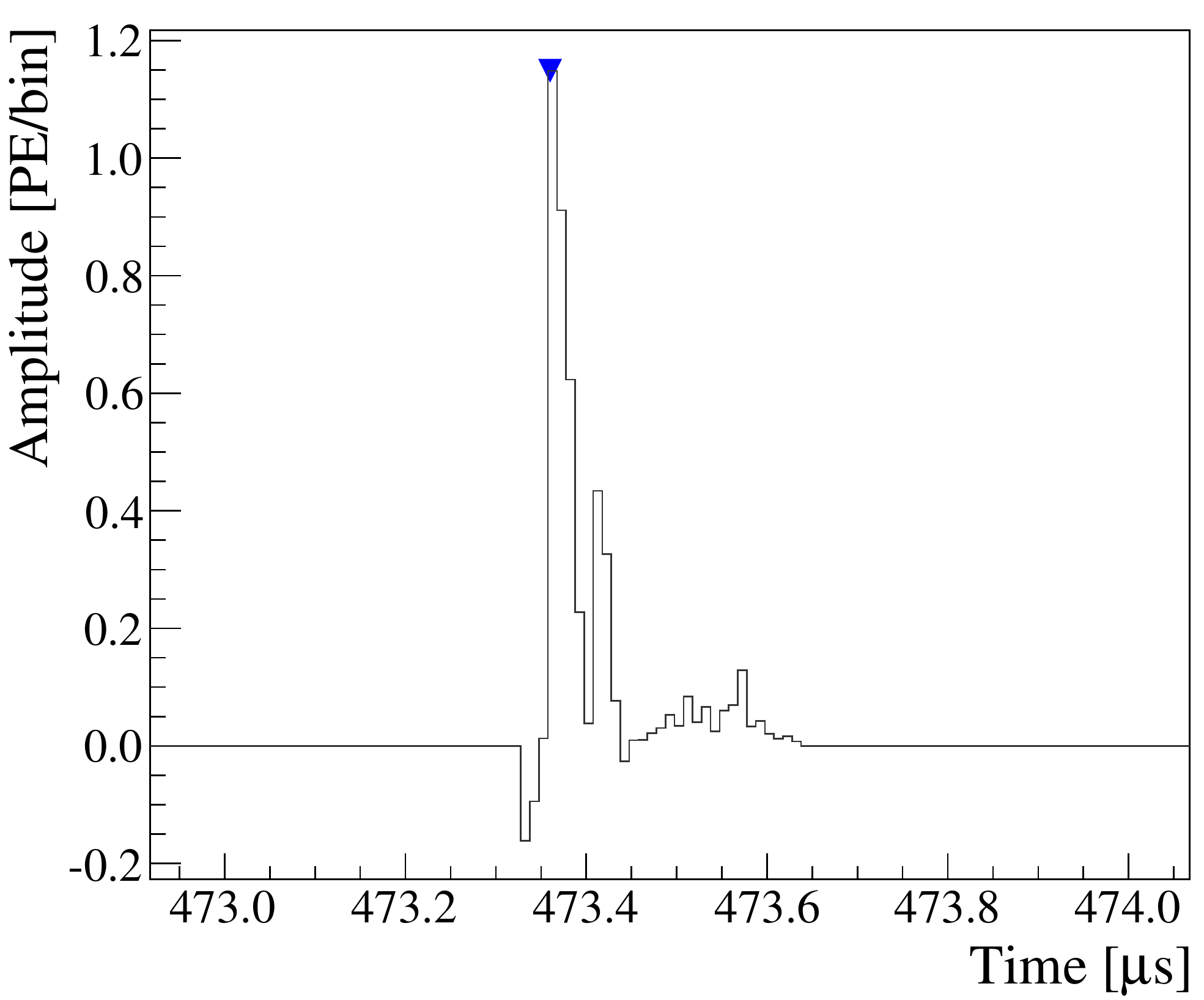}
\includegraphics*[width=0.38\textwidth]{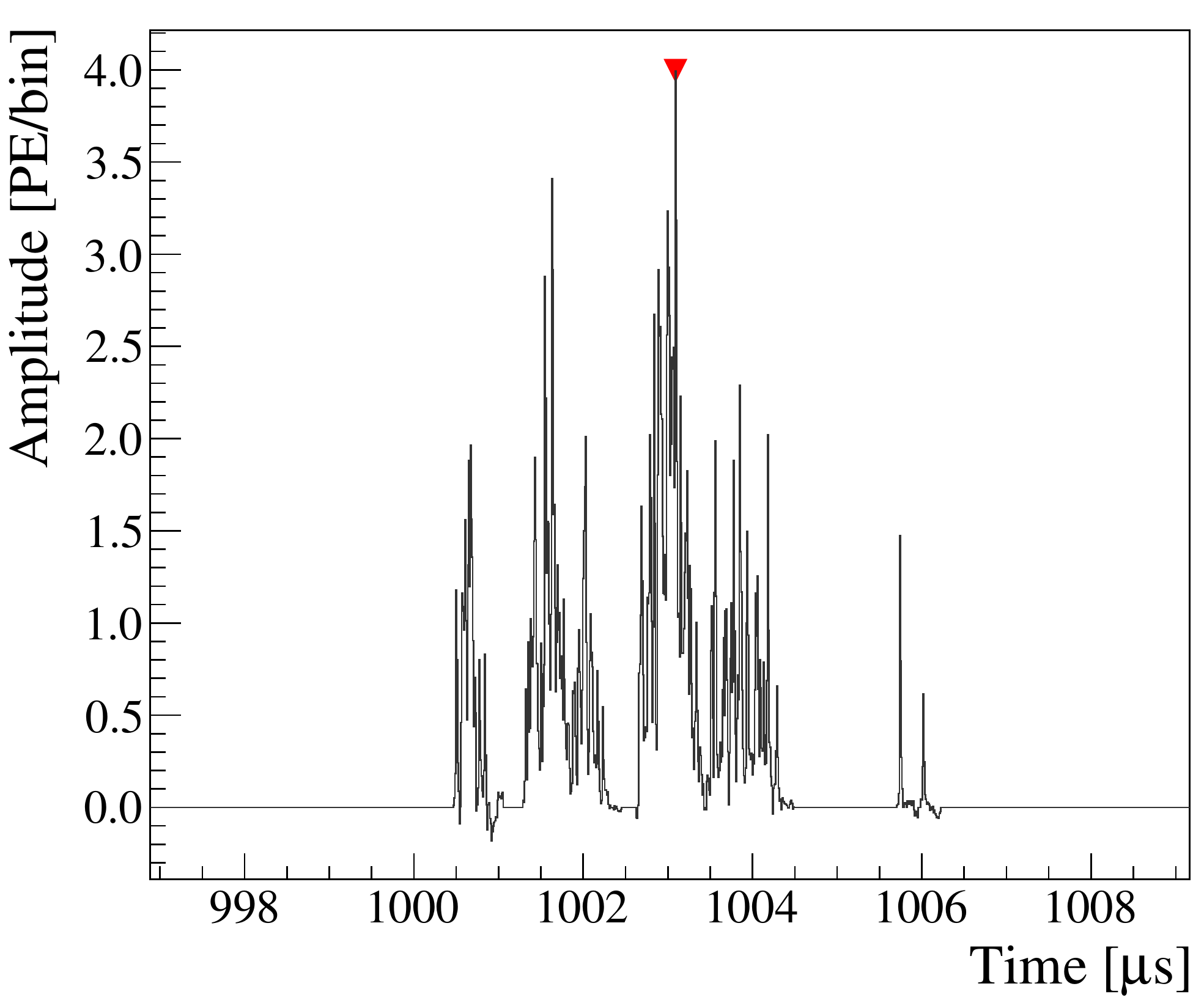}
\caption{Low-energy single-scatter NR event recorded by XENON1T during a $^{241}$AmBe calibration run. All panels show the sum of all TPC PMT waveforms. An S1~signal of 4.3\,PE (blue marker) is followed by an S2~signal of 250\,PE (red), where the quoted numbers do not yet take into account corrections. The drift time of 529.7\,$\mu$s corresponds to a depth of $Z=-75.9$\,cm. The lower panels focus on the main S1 and S2~peak, respectively. The smaller signals on the event trace are uncorrelated pulses from PMT dark counts, which are only seen by single PMT channels.
\label{fig::event}}
\end{figure*}

\begin{figure*}[p!]
\centering
\includegraphics*[width=0.48\textwidth]{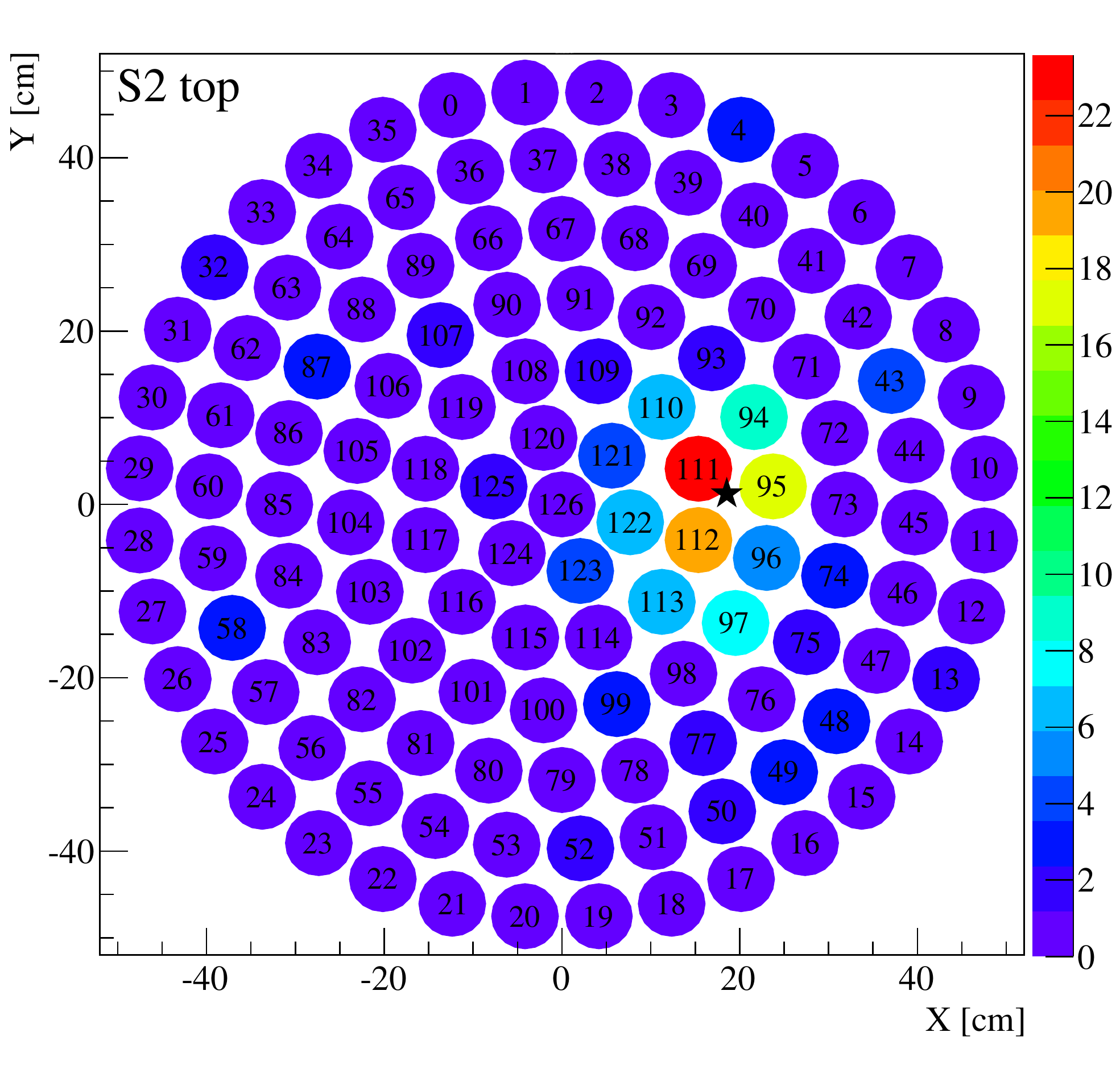}
\includegraphics*[width=0.48\textwidth]{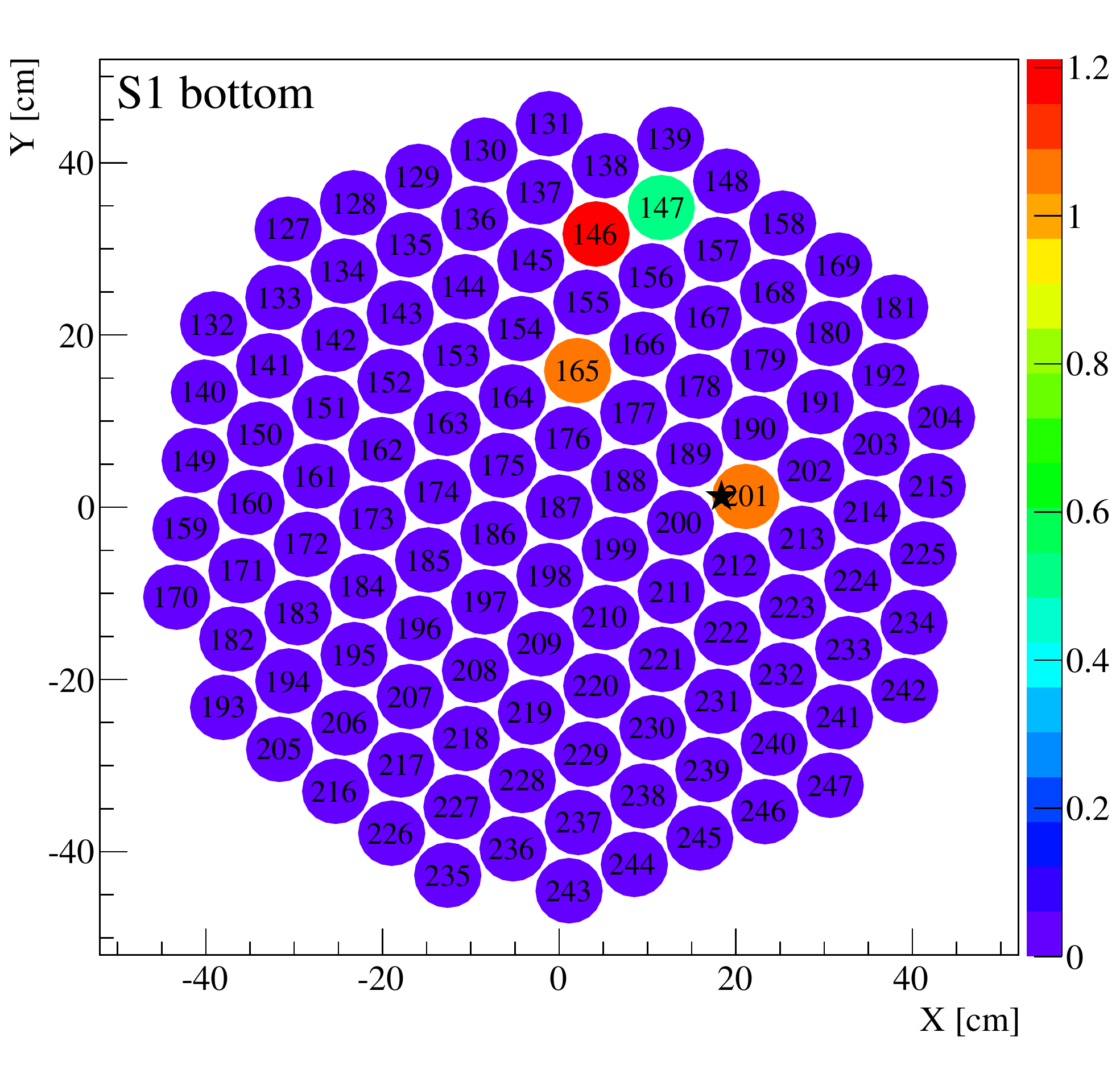}
\caption{Distribution of the S2 signal on the top (left) and the S1 signal on the bottom PMT array (right) of the event shown in figure~\ref{fig::event}. The color scales are given in uncorrected PE. The S2~signal is very localized and allows for the reconstruction of the lateral ($XY$) event position (star-shaped marker). This low-energy single-scatter NR event is located well inside in a fiducial volume containing 1\,t of LXe. 
\label{fig::hitpattern}}
\end{figure*}

The muon detection efficiency under these trigger conditions was obtained by means of a Monte Carlo simulation, taking into account all relevant signal creation and detection aspects. The muon veto is able to tag 99.5\% of the muons passing through the shield. Using the emission characteristics of Cherenkov light, the distribution of the signal arrival times on the 84~PMTs can be used to reconstruct the muon track through the water (see figure~\ref{fig::mvevent}). The same simulation was also used to assess the efficiency to tag muon-induced neutrons from a shower generated by a muon interaction outside of the water shield, which is 43\%. It is planned to reduce the trigger to a coincidence level of $N_\textnormal{\scriptsize pmt}=5$ in the future to achieve the design efficiency of tagging $>$70\% of the neutrons from a muon-induced shower~\cite{ref::xe1t_mv}.

The distribution of time differences between TPC events and their closest muon veto trigger during \srzero \ shows the presence of tight time coincidences ($<$1\,$\mu$s) which appear on top of accidental background. This population originates from simultaneous events in the two detectors and can be vetoed with a simple time coincidence cut.  Since the muon veto trigger rate is small, the rate of accidental coincidences between the two systems ($R_\textnormal{\scriptsize MV}=0.35$\,Hz, $R_\textnormal{\scriptsize TPC}\sim5$\,Hz) is small during a dark matter run. This leads to a negligible loss of live time even if only the muon veto trigger information is used to reject TPC events, without further analysis of the digitized muon veto PMT data. In \srzero, no low-energy event in the TPC was rejected because of a muon veto coincidence~\cite{ref::xe1t_sr0}.

%%%%%%%%%%%%%%%%%%%%%%%%%%%%%%%%%%%%%%%%%%%%%%%%%%%%%%%%%%%%%%%%%%%%%%
\subsection{Event Reconstruction} 
\label{sec::pax}

An example of raw TPC event data (``waveform'') from the DAQ system (see section~\ref{sec::daq}) is shown in figures~\ref{fig::event} and~\ref{fig::hitpattern} for a low-energy single-scatter NR. Physical quantities such as  signal area and position are reconstructed from the raw data by the custom-developed PAX data processor. PAX operates in five stages: (1) identify ``hits'' due to digitized photon signals in each PMT channel individually, (2) group hits into clusters, (3) compute properties for the clusters, (4) classify them as S1 or S2~candidates, and finally (5) compute physical quantities of interest for possible S1-S2 pairs. These stages are explained briefly below.

First, the processor finds hits in individual channels. For each ADC waveform segment (``pulse'', see section~\ref{sec::daq}), it computes the local baseline and noise level using samples before the self-triggering peak. Next, it searches the pulse for hits, defined as intervals above a threshold equal to or (for high-noise channels) higher than the digitization threshold, extended left by 30\,ns and right by 200\,ns to capture the entire photoelectron signal area.

Second, temporally nearby hits from all channels are clustered. The hits are initially split into groups separated by 2\,$\mu$s or more. Next,  these groups are split at local minima in the gain-corrected sum waveform using information from all identified hits, with significantly higher maxima on either side. Finally, we run a recursive algorithm minimizing the intra-cluster variance, up to a level calibrated on simulated waveforms.

Third, the PAX processor calculates properties for each hit cluster, such as total area, amplitude (in the gain-corrected sum waveform) and various measures of pulse width. The lateral $XY$-position of the cluster is computed by a likelihood maximizer that works with hitpatterns generated by optical Monte Carlo simulations. The reconstruction algorithm uses the results from simpler methods (maximum PMT, area-weighted mean, etc.) as an initial seed. It was shown in Monte-Carlo studies to have a radial resolution of $\Delta R<2$~cm. In addition the algorithm produces confidence contours on the reconstructed position.

Fourth, hit clusters are classified as S1, S2 or ``unknown'' peaks. A cluster is identified as an S1 if the rise-time of the signal is between 70~ns and 100~ns, and at least 3~PMTs contribute within 50\,ns. 
Slower rising clusters are classified as S2~signals if they are observed by at least 4~separate PMTs.

Finally, the processor considers every possible pairing of the largest S1 and S2~candidates. For each pair, the 3-dimensional position of the original interaction is reconstructed using the $XY$-position of the S2 signal and the $Z$-position from the time between the maxima of the~S1 and S2~peaks (``drift time''). The electron drift velocity is directly measured from the drift time distribution in the detector and the known TPC length. Using the position information, the signals are now corrected for spatial effects: primarily, light collection efficiency for~S1 and electron loss on impurities for~S2 (see section~\ref{sec::targetpur}). 

The muon veto raw data are time-synchronized to the TPC. They are searched for pulses from particle interactions using the same program package, however, in a slightly modified form since only prompt Cherenkov light pulses are expected to be detected.

%%%%%%%%%%%%%%%%%%%%%%%%%%%%%%%%%%%%%%%%%%%%%%%%%%%%%%%%%%%%%%%%%%%%%%
\subsection{Photomultiplier Performance}

The continuous long-term operation of the R11410-21 PMTs in XENON1T started in April~2016, when they were cooled-down in the TPC at a rate of 6\,mK/min. The PMTs have remained cold and immersed in LXe for more than 16~months and the majority of tubes showed a stable performance. Instead of aiming for the most uniform gain distribution, the individual bias voltages were chosen to maximize the PMT's individual single photoelectron (SPE) acceptance. For every PMT the highest stable operation voltage was selected, without exceeding a maximum of $-1550$\,V and a maximum gain of $5 \times 10^6$, to minimize light emission and elevated dark count rates. This optimizes the detector response to low signals and improves the S1~threshold. Individual gain variations are corrected by the data processor (see section~\ref{sec::pax}). 

The SPE acceptance is defined as the fraction of the SPE peak above the hardware ADC~digitization threshold (see section~\ref{sec::daq}), which is set to 15~ADC units ($=$2.06\,mV) for channels with the lowest electronic noise. At the start of \srzero~the SPE acceptance was around 89\% increasing to 93\% after optimization of the thresholds.

The PMT gains are regularly measured by stimulating the emission of SPEs by blue LED light. A new model-independent approach is employed~\cite{Saldanha:2016mkn} which allows extracting PMT parameters such as gain and occupancy without making assumptions on the underlying SPE distribution. The average gain is $2.6 \times 10^{7}$ when the additional $\times$10~amplification stage is taken into account. The gains are distributed between (2.0-5.0)\,$\times$\,$10^{7}$; the distribution has a rather wide standard deviation of $1.5 \times 10^7$. All gains were stable in time within 2\%, reflecting the uncertainty of the calibration method.

After installation in XENON1T, the PMT's average dark count rate decreased from $\sim$40\,Hz measured during the characterization campaign in gaseous $N_2$ at $-$100$^\circ$C~\cite{ref::r11410_character} to $\sim$12\,Hz and $\sim$24\,Hz for the top and bottom PMT arrays, respectively. The difference between the arrays is explained by the contribution of a larger fraction of LXe scintillation events to the dark count rate, which cannot be distinguished from ``real'' PMT dark counts. The overall reduction, likely thanks to the lower environmental radioactivity, is important to minimize the probability of accidental coincidences of uncorrelated dark count pulses, mimicking a low-energy S1~signal.

Even though PMTs indicating a loss of vacuum by the presence of lines from xenon ions (Xe$^+$, Xe$^{++}$) in their afterpulse spectra~\cite{ref::pmtpaper,ref::r11410_character} were not installed in the XENON1T PMT arrays, some tubes have developed new leaks during the operation in the cryogenic environment. The afterpulse spectra are thus investigated regularly to identify such tubes and to monitor the leak's evolution with time. The PMTs remain fully operational if the loss of vacuum is not too large. 28~PMTs showed a Xe-induced afterpulse rate of $r_\textnormal{\scriptsize Xe}>1$\%. Reducing their bias voltage helps to improve their performance, however, once their afterpulse rate becomes too large they have to be turned off. The PMTs with $r_\textnormal{\scriptsize Xe}\lesssim 1$\% were operated normally without a negative impact on the data analysis. Tubes with an identified leak showed a slight increase of the afterpulse rate of  $\Delta r_\textnormal{\scriptsize Xe}<0.1$\%/month.

During \srzero, a total of 27~R11410 PMTs (11~on the top and 16~on the bottom array) were switched off, corresponding to a loss of 11\%~of the channels. While the issues with 6~of the tubes are related to cabling and bad connections, the majority of the non-operational PMTs shows leak-related problems. The observed symptoms are PMT trips at rather low bias voltages, a high signal rate at the SPE level  and a ``flashing behavior''. The latter is characterized by a sudden increase of the PMT's trigger rate which lasts for a few minutes. The affected PMTs showed a high afterpulse rate, a clear indication for a leak, and seem to emit light during these periods as the rate of neighboring channels and of channels in the opposing array also increased. In many cases, these flashes appear to be triggered by high-energy depositions (e.g., from muons). Thanks to the large number of PMTs installed inside the XENON1T TPC, the impact of the missing channels on fiducialization is minor.

%%%%%%%%%%%%%%%%%%%%%%%%%%%%%%%%%%%%%%%%%%%%%%%%%%%%%%%%%%%%%%%%%%%%%%
\subsection{Target Purification} 
\label{sec::targetpur}

\begin{figure}[b!]
\includegraphics*[width=0.48\textwidth]{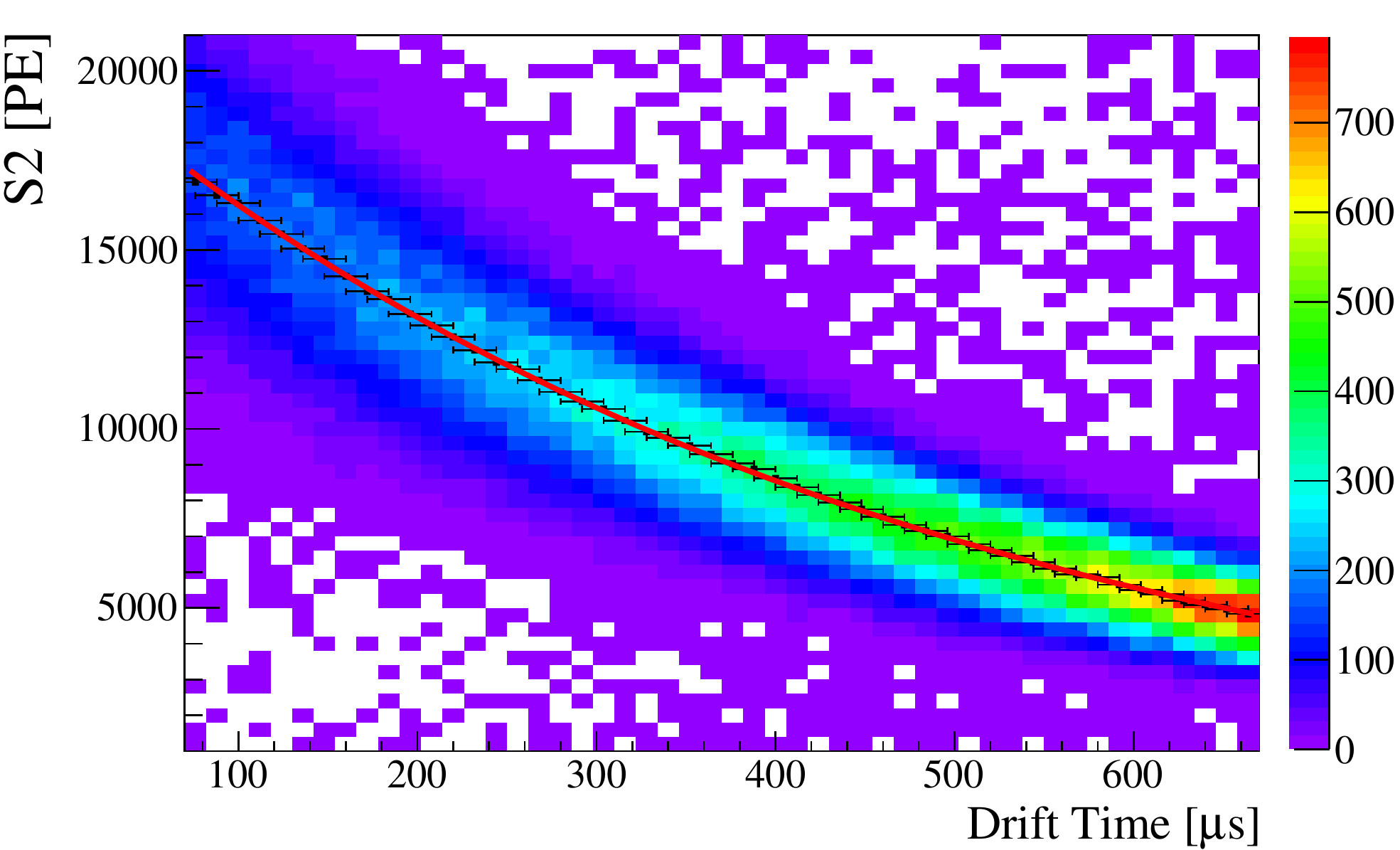}
\caption{The loss of electrons due to the presence of electronegative impurities in the LXe is measured via the drift time-dependent decrease of the charge signal size, e.g., from the \monoenergetic \ 32.1\,keV line from $^{83m}$Kr. The example shown here corresponds to an electron lifetime $\tau_e = (467 \pm 5)$\,$\mu$s, derived from the exponential fit (red line).  \label{fig::lifetime}}
\end{figure}

The loss of ionization electrons is caused by their attachment to electronegative impurities in the LXe target (H$_2$O, O$_2$, etc.) and it is described by the finite electron lifetime $\tau_e$, which thus serves as a measurement of the target purity. An initial charge signal of size S2$_0$ is exponentially reduced depending on the drift time $t$ between the interaction point and the liquid-gas interface:
\begin{equation}
\textnormal{S2}(t) = \textnormal{S2}_0 \ \exp(- t / \tau_e) \textnormal{.} 
\end{equation}
This drift-time dependent effect is the most important charge signal correction and is applied to every event. To compensate for outgassing from materials and to maximize $\tau_e$, the LXe target is constantly purified (see Sect.~\ref{sec::pur}). The electron lifetime $\tau_e$ is regularly measured by characterizing the signal loss of \monoenergetic \ charge signals (e.g., full absorption peaks) across the TPC. Figure~\ref{fig::lifetime} shows such a measurement using the \monoenergetic \ conversion electron line at 32.1\,keV from meta\-stable $^{83m}$Kr, which is very well described by an exponential function.

\begin{figure}[b!]
\includegraphics*[width=0.48\textwidth]{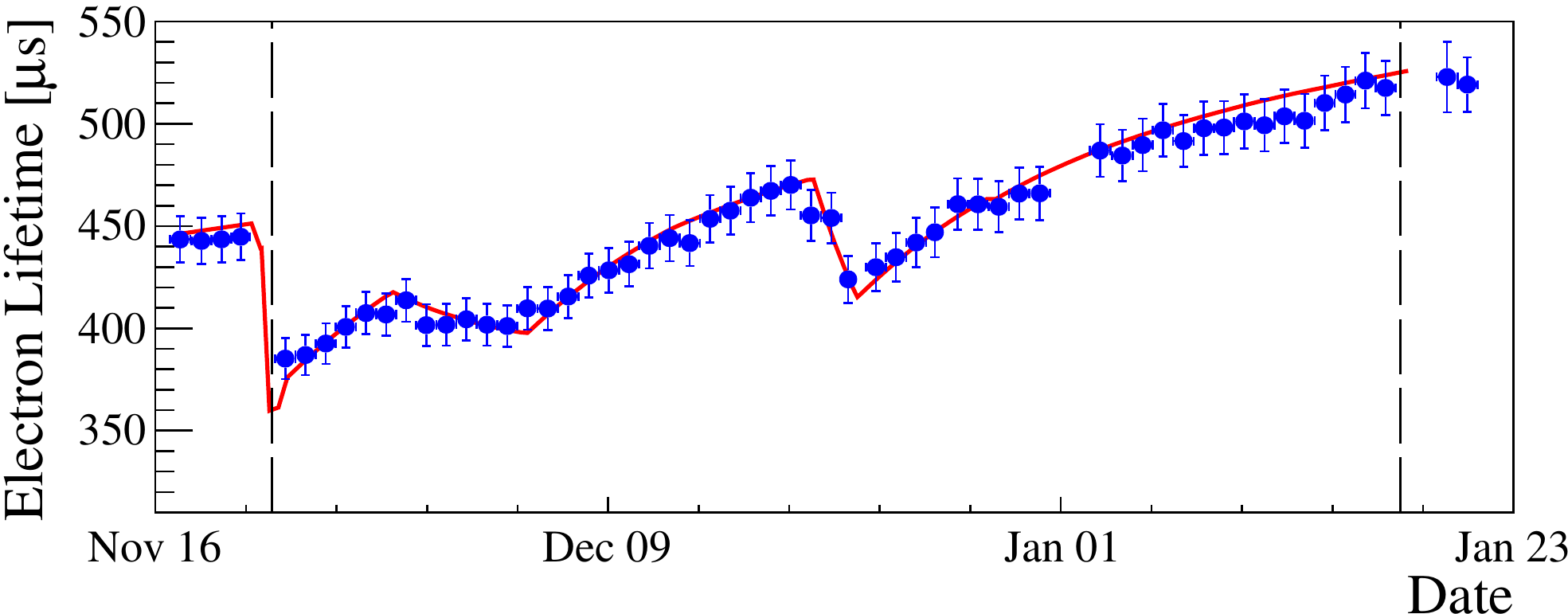}  
\caption{Evolution of the electron lifetime over \srzero, which is indicated by the vertical lines.  The two step-like decreases are well-understood and related to detector operation (e.g., online $^{222}$Rn distillation which started on December~19). The model (red line) describes the data well.  \label{fig::lifetime_evol}}
\end{figure}

Figure~\ref{fig::lifetime_evol} shows the electron lifetime evolution over a period of approximately two months, during \srzero~data acquisition. It was measured using \monoenergetic \ $\alpha$ peaks from the $^{222}$Rn decay chain observed in background data. During \srzero, $\tau_e$ varied between 350\,$\mu$s and 500\,$\mu$s, with an average of 452\,$\mu$s. The purity was limited by detector outgassing and a purification flow of 50~slpm ($\approx 425$ kg/d), above which the QDrive gas pumps could not operate reliably given the flow impedance in the circuit.

A few sudden decreases of the electron lifetime are visible in figure~\ref{fig::lifetime_evol}. They are all related to the detector operation conditions and can be modeled as shown by the fit. The decrease after December~19, 2016, for example, was caused by starting the online removal of radon from the Xe~target, which required re-routing part of the purification gas flow (see also figure~\ref{fig::sc_stability}).

In principle, also the primary scintillation signal~S1 could be affected by the LXe purity via light absorption on impurities (mainly H$_2$O). However, even in the very first measurements during detector commissioning, the attenuation length of the scintillation light of $>$10\,m was much larger than the detector dimensions.

%%%%%%%%%%%%%%%%%%%%%%%%%%%%%%%%%%%%%%%%%%%%%%%%%%%%%%%%%%%%%%%%%%%%%%
\subsection{Light and Charge Measurements in the TPC} 
\label{sec::tpcperformance}

A drift field of 0.125\,kV/cm in the TPC was established by biasing the cathode with $-$12\,kV and keeping the gate electrode at ground potential. The maximal drift time at this field is 673\,$\mu$s for events occurring at the cathode. With the known TPC length, a drift velocity of 1.44\,mm/$\mu$s was calculated. The anode was biased with $+$4.0\,kV, leading to an 8.1\,kV/cm extraction field across the liquid-gas interface, which was placed right between gate and anode electrodes, 2.5\,mm above the gate. The top and bottom screening electrodes were biased with $-$1.55\,kV to realize a field-free region in front of the PMT photocathodes.

Due to the lower drift field compared to the initial design, boundary effects to the field become relatively more important, especially close to the TPC walls and close to the cathode. The field lines at large radii are no longer straight but bend toward smaller radii (see figure~\ref{fig::tpc_efield}). This leads to an inward-bias of the reconstructed event positions but ensures that no charge signal is lost. An ($R,Z$)-dependent field correction map was constructed using an axisymmetric 3D~finite element simulation of the realized geometry. The map was verified by means of calibration data ($^{83m}$Kr, $^{220}$Rn) and using $\alpha$ events which clearly define the PTFE TPC walls. In a central, cylindrical fiducial volume of 1\,t, the radial difference between reconstructed and real interaction positions was determined to be below 10\,mm for most of the volume, but it can increase to $\sim$50\,mm for high radii and depths.

The TPC was calibrated by means of various internal and external sources to measure its response to light and charge signals. In \srzero, a $^{241}$AmBe source was used to calibrate the NR response; the neutron generator was successfully employed in science run~1.

\begin{figure}[b!]
\includegraphics*[width=0.48\textwidth]{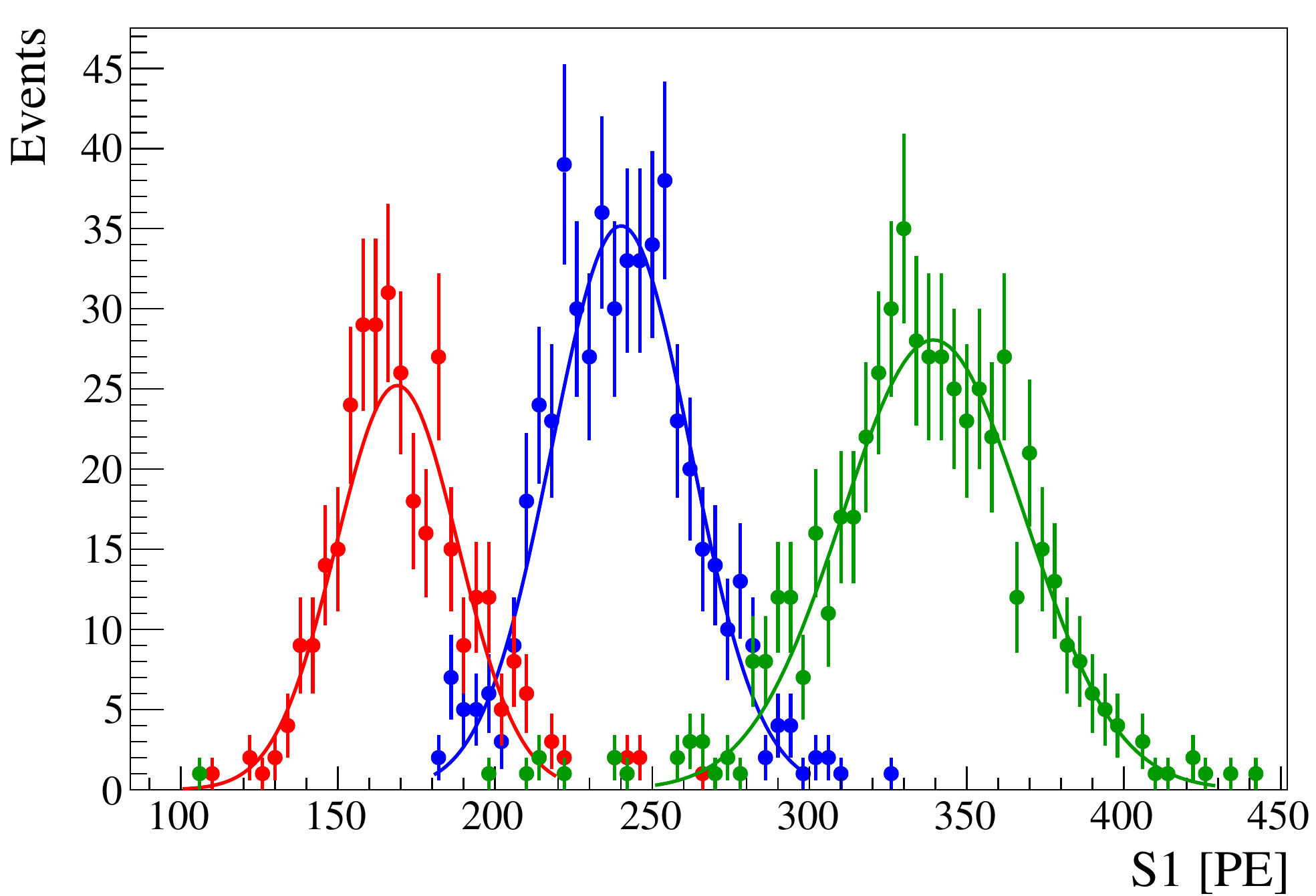}   
\caption{Primary scintillation light (S1) from $^{83m}$Kr 32.1\,keV conversion electron events in detector regions close to the central axis ($R=8\ldots16$\,cm) but at different heights: just below the gate (red), in the TPC center (blue) and above the cathode (green). These spectra were used to construct the correction map shown in figure~\ref{fig::lce}.\label{fig::light_spectra}}
\end{figure}

\begin{figure}[b!]
\centering
\includegraphics*[width=0.35\textwidth]{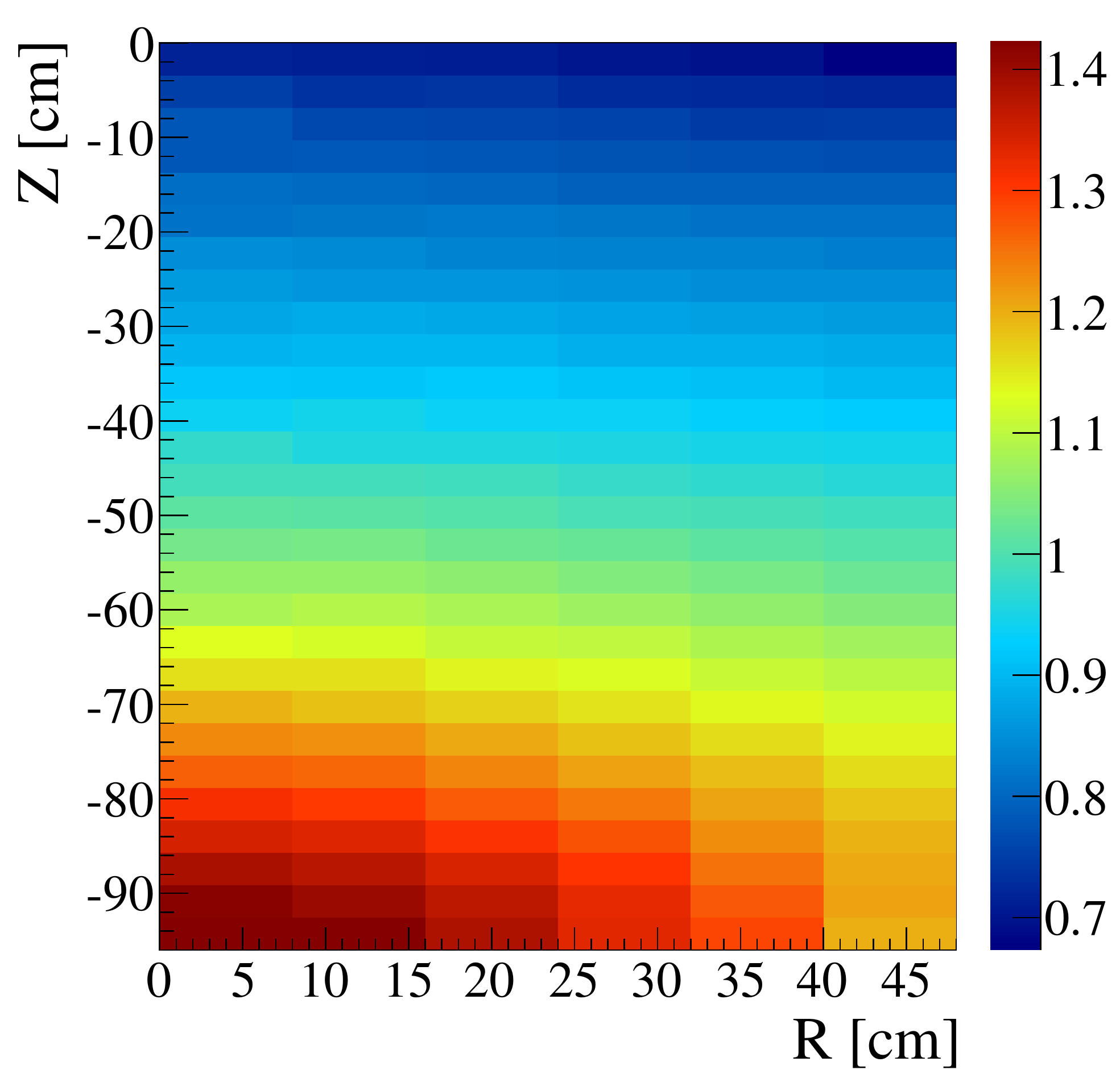}   
\caption{Azimuthally averaged light collection efficiency relative to the mean light yield across the XENON1T TPC. The variation is mainly caused by solid-angle effects and light-reflection. The efficiency is maximal right above the center of the bottom PMT array and minimal below the outer ring of the top PMT array. \label{fig::lce}}
\end{figure}

\paragraph{Light Signal} 
The light collection efficiency in the active TPC volume is not uniform but affected by factors such as the solid-angle coverage of the PMTs and the average number of reflections before a photon hits a photocathode. In order to correct for this effect, the \monoenergetic \ light signal induced by 32.1\,keV conversion electrons from $^{83m}$Kr was measured in discrete ($R,Z$)-regions, see figure~\ref{fig::light_spectra}. The mean of the individual distributions was used to construct a correction map relative to the mean light yield in the TPC (figure~\ref{fig::lce}). The light collection efficiency varies by a factor $\sim$2 across the TPC, with the maximum being in the TPC center, right above the cathode. The minimum is below the outermost ring of top PMTs. A small dependence on the azimuth-angle~$\phi$ is taken into account in the correction function used by the peak processor.

\begin{figure}[b!]
\centering
\includegraphics*[width=0.4\textwidth]{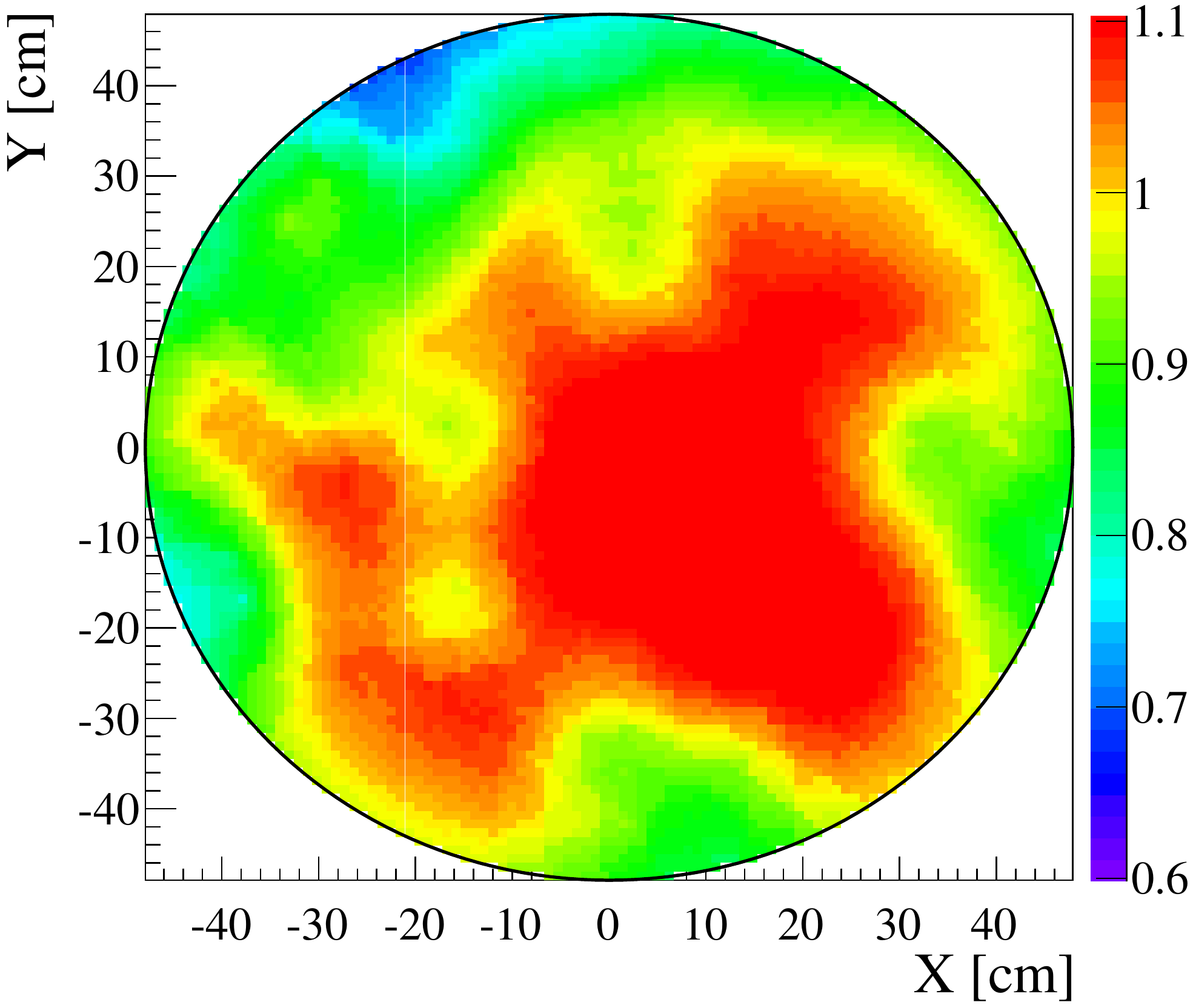}  
\includegraphics*[width=0.4\textwidth]{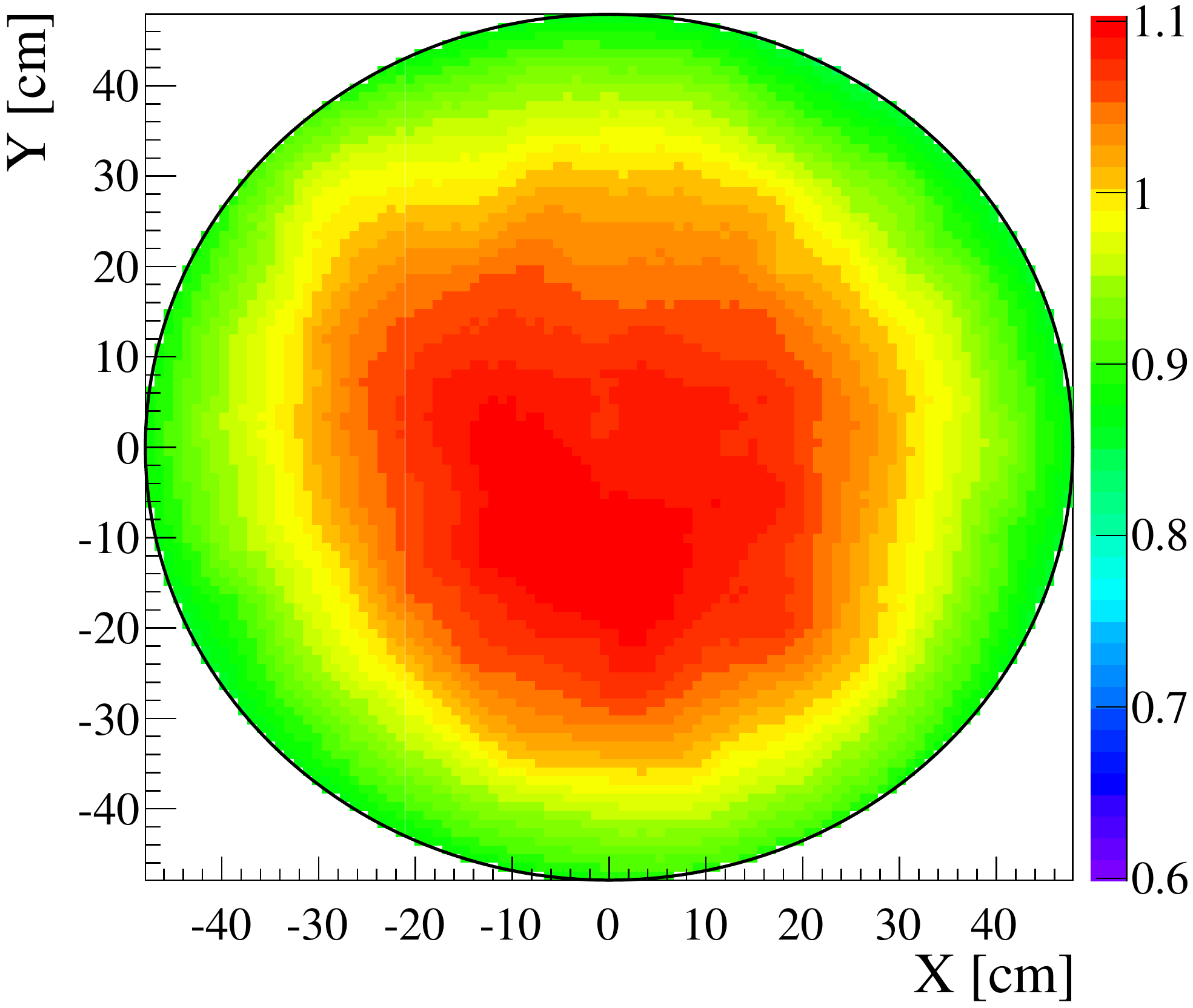}  
\caption{Relative S2-correction maps compensating the non-uniform charge response of the top and bottom PMT arrays. The response of the top array is more affected by local non-uniformities such as non-operational PMTs. The response of the bottom array is significantly more uniform.\label{fig::s2maps}}
\end{figure}

%%%%%%%%%%%%%%%%%%%%%%%%%%%%%%%%%%%%%%%%%%%%%%%%%
\paragraph{Charge Signal}

The measurement of the charge signal is also affected by solid-angle and other detector-related effects. The proportional scintillation signal~S2 is generated in a well-defined plane between the liquid-gas interface and the anode electrode, about 7.5\,cm below the PMTs of the top array. About half of the light is thus observed by a few top PMTs just above the S2~production region, while the other half is rather uniformly distributed over the bottom PMT array. In order to reconstruct the number of electrons producing the signal, S2 correction maps are required, see figure~\ref{fig::s2maps}. These were derived from the combined S2~signal (41.5\,keV) from $^{83m}$Kr since the short time separation between the two S2~peaks (half-life $T_{1/2}=154$\,ns of the intermediate state~\cite{Manalaysay:2009yq}) makes it challenging to separate the two contributions and reduces the size of the data sample. The response of the top array shows local variations at the (10-15)\,\% level, which are mainly caused by non-functional PMTs. A slight increase of S2~signal is visible towards the center, which is due to the sagging of the anode electrode. At the location of lowest S2~response ($X=-20$, $Y=40$), two neighbouring PMTs are non-functioning. 

The S2~response of the bottom PMT array is much more homogeneous. It can be mainly explained by solid-angle coverage and does not show significant local variations. The size of the S2~correction is thus less affected by the uncertainty in the reconstructed event position and leads, for example, to a slightly better energy resolution. For this reason, only the S2~signal from the bottom array, S2$_\textnormal{\scriptsize b}$, was used as an energy estimator for the analysis of \srzero~\cite{ref::xe1t_sr0}.

%%%%%%%%%%%%%%%%%%%%%%%%%%%%%%%%%%%%%%%%%%%%%%%%%
\paragraph{Light and Charge Yield}

\begin{figure}[b!]
\includegraphics*[width=0.48\textwidth]{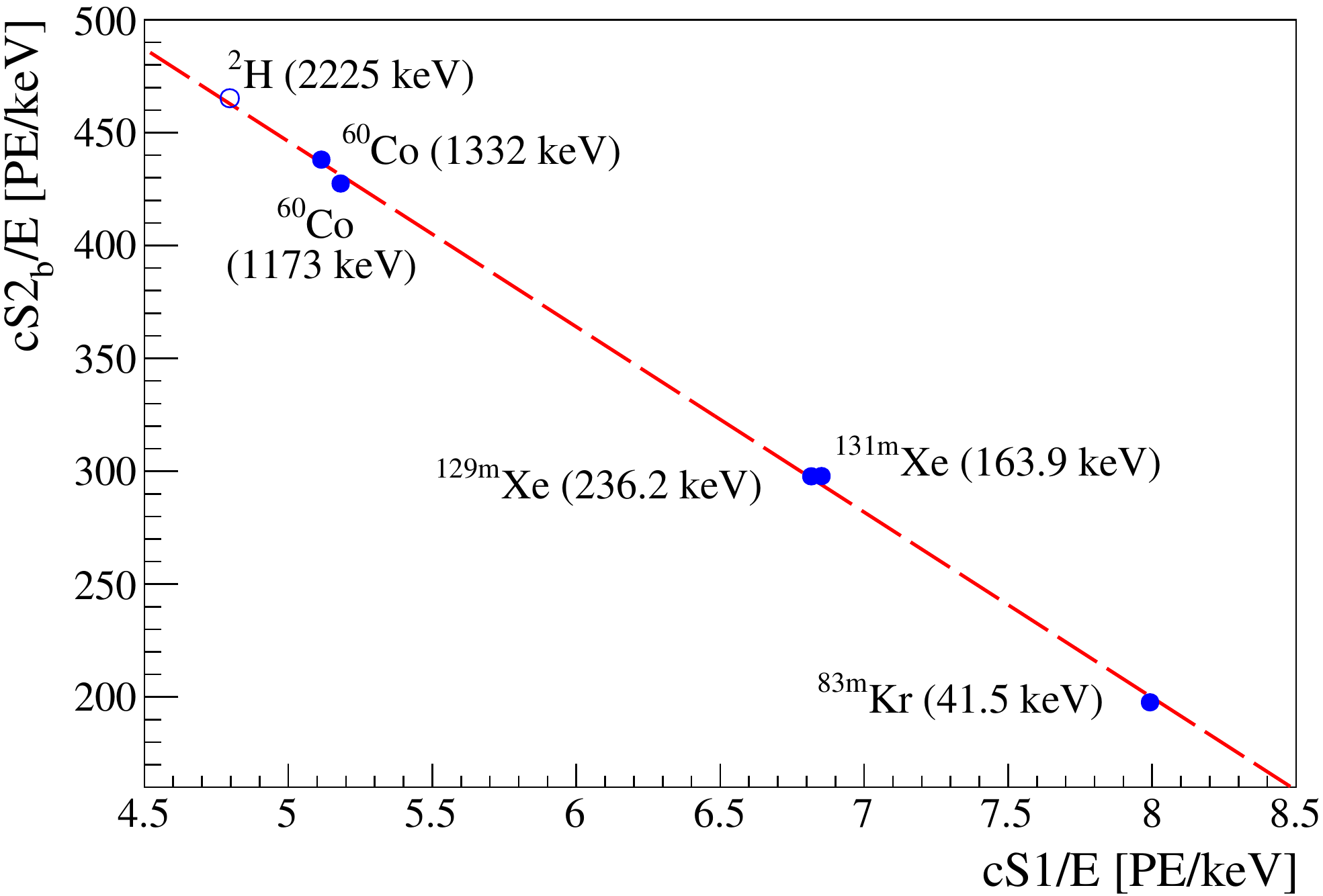} 
\caption{The scintillation (cS1) and ionization signals (cS2) for various \monoenergetic \ peaks, corrected for position dependent effects and normalized to the line energy, show the expected anti-correlated behavior. The fit to the data allows the extraction of the primary (in PE/$\gamma$) and secondary scintillation gain (in PE/e$^-$). The signal at 2.2\,MeV is due to de-excitation $\gamma$ rays from neutron
capture on hydrogen~(${}^1$H$(n,\gamma){}^2$H). It is not used for the fit, but demonstrates that the detector response is well known over a large energy range. 
\label{fig::doke}}
\end{figure}

The parameters describing the detector's efficiency to detect light and charge signals are the primary scintillation gain $g_1$\,=\,cS1/$n_\gamma$ and the secondary scintillation gain $g_{2b}$\,=\,cS2$_\textnormal{\scriptsize b}$/$n_e$, where the observables cS1 and cS2$_\textnormal{\scriptsize b}$ are corrected for position-dependent effects. Almost all electronic recoil energy $E$ is used
for the production of photons ($\gamma$) and electrons ($e$),
\begin{equation}\label{eq::doke}
E=(n_\gamma + n_e) \times W = \left( \frac{\textnormal{cS1}}{g_1} + \frac{\textnormal{cS2}_\textnormal{\scriptsize b}}{g_{2b}} 
\right) \times W \textnormal{,}
\end{equation}
where $W$\,$=$\,13.7\,eV is the average energy required to produce one electron-ion pair or to excite one Xe~atom~\cite{ref::nestW}. The two observables are anti-correlated, which can be exploited to improve the energy resolution for ER signals. Figure~\ref{fig::doke} shows the determination of $g_1$ and $g_{2b}$ using several \monoenergetic \ peaks, which fall on a straight line once the observables are normalized to the peak energy. Re-arranging equation~(\ref{eq::doke}) allows for the extraction of $g_1= (0.144 \pm 0.007)$\,PE/$\gamma$ and $g_{2b}=(11.5 \pm 0.8)$\,PE/e$^-$ from a fit. The uncertainty combines statistical and systematic uncertainties. These parameters neither depend on the line's energy nor the TPC drift field. Taking into account the emission of two photoelectrons by one photon~\cite{Faham2015}, the measured $g_1$-value corresponds to a photon detection efficiency of $(12.5\pm0.6)$\%, which is consistent with the design value~\cite{ref::xe1t_reach}.
The secondary scintillation gain $g_{2b}$ from the fit is in agreement with the one obtained from describing the charge spectrum at lowest energy, acquired in the tails of regular S2~peaks, with a Gaussian with a mean of $(11.7 \pm 0.3)$\,PE/e$^-$. $g_{2b}$ is related to the electron extraction efficiency, which is calculated to be 96\%. 

\begin{figure}[t!]
\includegraphics*[width=0.48\textwidth]{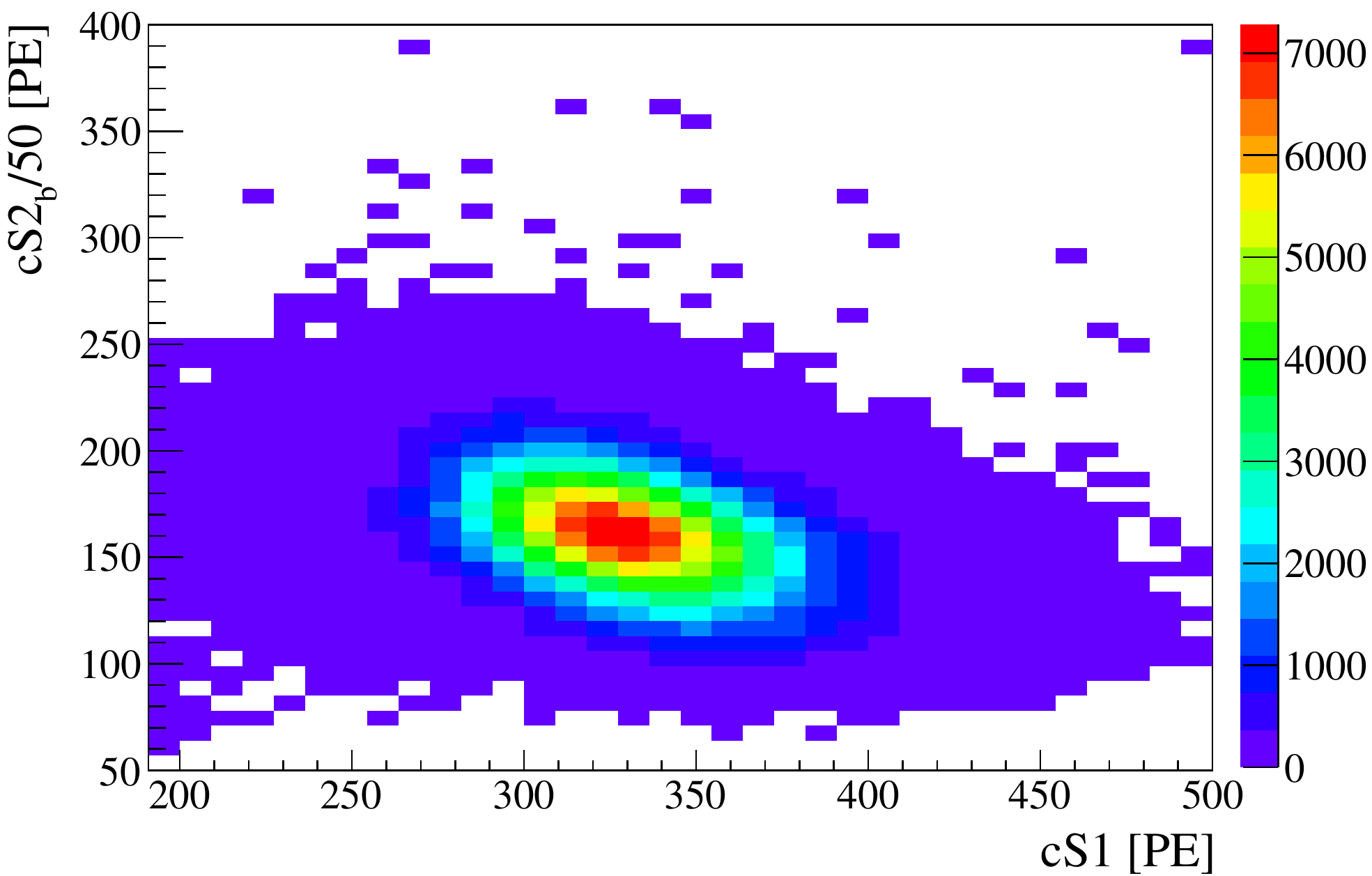}  
\caption{The position-corrected S1 and bottom S2 signals of the $^{83m}$Kr isomer decay allows for the determination of the light and carge yields at 41.5\,keV. The $^{83m}$Kr decay produces two consecutive conversion electrons which are combined for this analysis.
\label{fig::kr83m}}
\end{figure}

Light and charge yield are defined as the number of photoelectrons measured at a reference energy and operational drift field, which is 0.125\,kV/cm for the XENON1T \srzero. Figure~\ref{fig::kr83m} shows the corrected S1 and bottom S2 signals for the 41.5\,keV line from $^{83m}$Kr. Describing both signals by Gaussians leads to a light yield of ($8.02 \pm 0.06$)\,PE/keV and a charge yield of  $(198.3 \pm 2.3)$\,PE/keV at 41.5\,keV. The anti-correlation of both signals is clearly visible.

\begin{figure}[t!]
\includegraphics*[width=0.48\textwidth]{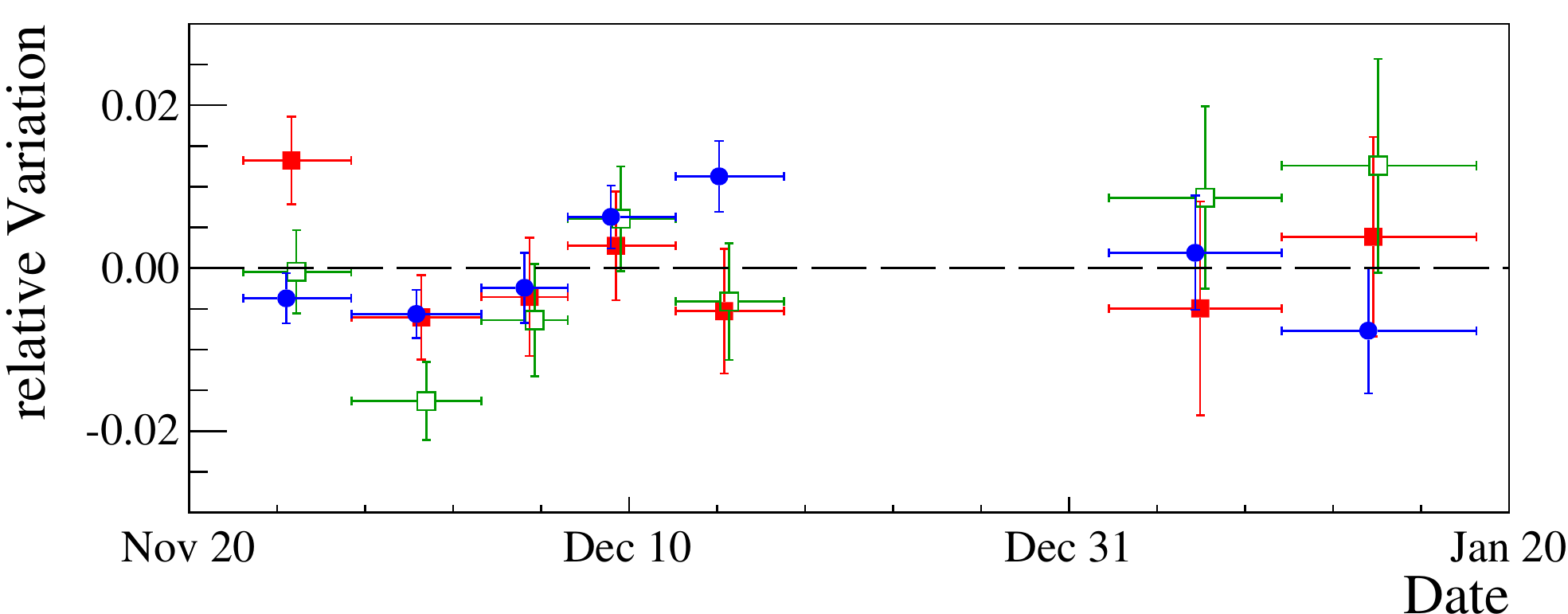}   
\caption{The light and charge response is stable with time as demonstrated using the 164\,keV line from activated $^{131m}$Xe. Shown are the relative variations of the light response (filled, blue points) and the charge response in various time intervals. The charge response is shown separately for both PMT arrays combined (red, filled squares) and the bottom array only (green, open squares). After applying all corrections, the response is stable to $<$2\,\%. \label{fig::stability}}
\end{figure}

As data from volume calibrations with $^{83m}$Kr were not available throughout \srzero, the 164\,keV line from $^{131m}$Xe was used to monitor the stability of the light and charge yields over time.  $^{131m}$Xe is produced in the target by activating xenon isotopes using neutrons from the $^{241}$AmBe calibration campaign. Light yield and charge yield (both PMT arrays and bottom array only) were stable at the $<$2\% level throughout the run as shown in figure~\ref{fig::stability}. As the $^{131m}$Xe activity decreases with time, less data were available at later times, which required larger binning. No data were available for a period of two weeks towards the end of December, when a low-energy bulk calibration with $^{220}$Rn prevented the identification of a clean $^{131m}$Xe peak.

%%%%%%%%%%%%%%%%%%%%%%%%%%%%%%%%%%%%%%%%%%%%%%%%%%%%%%%%%%%%%%%%%%%%%%
%%%%%%%%%%%%%%%%%%%%%%%%%%%%%%%%%%%%%%%%%%%%%%%%%%%%%%%%%%%%%%%%%%%%%%
%%%%%%%%%%%%%%%%%%%%%%%%%%%%%%%%%%%%%%%%%%%%%%%%%%%%%%%%%%%%%%%%%%%%%%
\section{Outlook}
\label{sec::outlook}

The XENON1T detector is currently operating in stable conditions underground at LNGS. With the release of the first results from the short \srzero~of only 34\,live-days, it demonstrated that it is the most sensitive dark matter search experiment for spin-independent WIMP-nucleon scattering for WIMP masses above 10\,GeV/$c^2$~\cite{ref::xe1t_sr0}. To fully exploit the physics potential of the experiment, a long science run with a livetime of approximately two years is necessary~\cite{ref::xe1t_reach}.

The next generation instrument, XENONnT, with a total mass of $\sim$8\,t of LXe and with 6\,t in the target, is already in the technical design phase. It will increase the sensitivity by another order of magnitude compared to XENON1T~\cite{ref::xe1t_reach} and will be able to confirm a WIMP detection in case XENON1T would see an excess of events. XENON1T will continue to acquire science data until the construction of all XENONnT components is finalised. We foresee a $\sim$6\,month interruption of XENON data-taking for the de-commissioning of XENON1T, and subsequent installation and commissioning 
of XENONnT.

%%%%%%%%%%%%%%%%%%%%%%%%%%%%%%%%%%%%%%%%%%%%%%%%%%%%%%%%%%%%%%%%%%%%%%
\begin{acknowledgements}
We would like to thank D.~Bar, R.~Berendes, G.~Bourichter, G.~Bucciarelli, S.~Felzer, S.~Form, B.~Gramlich, M.~Guerzoni, R.~H\"{a}nni, R.~Hofacker, G.~Korga, A.~Korporaal, P.~Martella, R.~Michinelli, G.~Panella, M.~Reissfelder, D.~Stefanik, G.~Tajiri, M.~Tobia and J.~Westermann for their contributions to our experiment. We gratefully acknowledge support from the National Science Foundation, Swiss National Science Foundation, Deutsche Forschungsgemeinschaft, Max Planck Gesellschaft, German Ministry for Education and Research, Netherlands Organisation for Scientific Research, Weizmann Institute of Science, I-CORE, Pazy-VATAT, Initial Training Network Invisibles (Marie Curie Actions, PITNGA-2011-289442), Fundacao para a Ciencia e a Tecnologia, Region des Pays de la Loire, Knut and Alice Wallenberg Foundation, Kavli Foundation, and Istituto Nazionale di Fisica Nucleare. We are grateful to Laboratori Nazionali del Gran Sasso for hosting and supporting the XENON project.  
\end{acknowledgements}

\end{document}